\documentclass[showpacs,pre,twocolumn]{revtex4}%
\usepackage{amsfonts}
\usepackage{amsmath}
\usepackage{amssymb}
\usepackage{graphicx}%
\providecommand{\U}[1]{\protect\rule{.1in}{.1in}}

\begin{document}
\title{Generalized Lyapunov Exponent and Transmission Statistics in One-dimensional
Gaussian Correlated Potentials}
\author{E. Gurevich and A. Iomin}
\affiliation{Department of Physics, Technion, Israel Institute of Technology, Haifa 32000, Israel}

\begin{abstract}
Distribution of the transmission coefficient $T$ of a long system with a
correlated Gaussian disorder is studied analytically and numerically in terms
of the generalized Lyapunov exponent (LE) and the cumulants of $\ln T$. The
effect of the disorder correlations on these quantities is considered in weak,
moderate and strong disorder for different models of correlation. Scaling
relations between the cumulants of $\ln T$ are obtained. The cumulants are
treated analytically within the semiclassical approximation in strong
disorder, and numerically for an arbitrary strength of the disorder. A small
correlation scale approximation is developed for calculation of the
generalized LE in a general correlated disorder. An essential effect of the
disorder correlations on the transmission statistics is found. In particular,
obtained relations between the cumulants and between them and the generalized
LE show that, beyond weak disorder, transmission fluctuations and deviation of
their distribution from the log-normal form (in a long but finite system) are
greatly enhanced due to the disorder correlations. Parametric dependence of
these effects upon the correlation scale is presented.

\end{abstract}
\maketitle

\section{Introduction}

Properties of stochastic systems can be significantly affected by the degree
of correlation of the external noise term (see, e.g., Ref. \cite{Hanggi-95}
and references therein). As an example of such phenomenon, we study
statistical properties of Anderson localization \cite{Anderson-58} in a linear
one-dimensional model%
\begin{equation}
-\frac{d^{2}\psi}{dx^{2}}+V(x)\psi=\epsilon\psi, \label{Schrodinger}%
\end{equation}
with correlated disorder potential $V(x)$. With a suitable change of notation,
equation (\ref{Schrodinger}) can describe a stationary problem either for a
quantum particle at energy $\epsilon$, or for classical scalar electromagnetic
or acoustic waves. Equation (\ref{Schrodinger}) appears also in many other
fields of physics. For example, with $x$ considered time, Eq.
(\ref{Schrodinger}) represents a random frequency oscillator, which is a
simple paradigm of a stochastic dynamical system (see Ref. \cite{Pikovsky-03}
for a more detailed discussion of this and other applications). In all these
instances, an important quantity is the Lyapunov exponent (LE) $\gamma
=\lim_{x\rightarrow\infty}\tilde{\gamma}\left(  x\right)  $, where
$\tilde{\gamma}\left(  x\right)  =\frac{1}{2x}\ln\left(  \left\vert
\psi\right\vert ^{2}+\left\vert \psi^{\prime}\right\vert ^{2}\right)  $ and
$\psi$ is a solution of an initial value problem. LE is a non-random quantity
independent of the specific realization of disorder $V\left(  x\right)  $
\cite{LGP-Introduction}. On the contrary, "local" LE $\tilde{\gamma}\left(
x\right)  $ is random and should be described statistically by distribution
$P\left(  \tilde{\gamma};x\right)  $. Distribution $P\left(  \tilde{\gamma
};x\right)  $ can be studied in terms of its cumulants (e.g. Ref.
\cite{Schomerus-02}) or using the generalized Lyapunov exponents
$\Lambda^{\left(  q\right)  }=\lim_{x\rightarrow\infty}\frac{1}{2qx}%
\ln\left\langle \left(  \left\vert \psi\right\vert ^{2}+\left\vert
\psi^{\prime}\right\vert ^{2}\right)  ^{q/2}\right\rangle $, where
$\left\langle \cdots\right\rangle $ denotes average over the disorder
realizations (e.g., Refs. \cite{Pikovsky-03},\cite{Paladin-87}).

LEs $\gamma$ and $\tilde{\gamma}\left(  x\right)  $ are intimately related to
the localization properties of one-dimensional disordered systems. The latter
was studied extensively and it is rigorously established that almost all
eigenstates in one-dimentional (1D) disordered systems are exponentially
localized under rather general conditions \cite{Kotani-82},\cite{Simon-83}%
,\cite{Kotani-87} (see also \cite{LGP-Introduction},\cite{Pastuf-Figotin} and
references therein). The inverse localization length of the eigenstates, as
well as the asymptotic decay rate $\lim_{L\rightarrow\infty}\left(  2L\right)
^{-1}\ln T^{-1}\left(  L\right)  $ of the transmission coefficient $T$ of the
system of length $L$, are equal \cite{LGP-Introduction} to the LE $\gamma$.
More generally, statistical properties of the quantity $\left(  2L\right)
^{-1}\ln T^{-1}\left(  L\right)  $ and of the introduced above local LE
$\tilde{\gamma}\left(  x=L\right)  $ become asymptotically equivalent for
large $L$ (see Sec. \ref{Sect: Model}). Therefore, in what follows, we discuss
the problem in terms of the transmission coefficient $T$, while the results
apply in a broader context of the local LE $\tilde{\gamma}$.

Transmission coefficient is a basic characteristics of the wave transport
through the non-uniform media. In the electronic systems, $T$ is related to
the dimensionless conductance by the Landauer formula \cite{Landauer-70}.
Measurement of the transmission coefficients is one of the ways to investigate
different aspects of Anderson localization experimentally. Some indications of
Anderson localization were observed with light \cite{Light-Loc}, microwaves
\cite{Micro-W-Loc}, cold atoms \cite{Aspect-1DBECexper-08} and ultrasound
\cite{Sound-Loc}. In most of these experiments, the correlation scale of the
randomness is comparable to the scattered wavelength and should be taken into
account to obtain a full quantitative description. In the experiments in Ref.
\cite{Aspect-1DBECexper-08}, localization of cold atoms was found by observing
localized density profiles, rather than transmission. Recently, possible
experiments on cold atom transmission through the disordered optical
potentials were discussed \cite{Ernst-10}. It is worth noting that in the
actively developing field of cold atoms, one of the ways to introduce disorder
is by using the laser speckle intensity patterns, which are highly correlated
\cite{Aspect-Speckle-06}. On the theoretical side, there is an increasing
interest in the disordered systems with correlated potentials. In particular,
in 1D systems, disorder correlations were found to have strong effect both on
the localization length \cite{LGP-Introduction},\cite{Israilev-99}%
,\cite{Tessiery-Israilev-01},\cite{Gurevich},\cite{Palencia-09}, on the
generalized LE \cite{Tessieri-02},\cite{Iomin-09} and on the transmission
statistics \cite{Titov-05},\cite{Deych-03} and, in special cases, even lead to
the appearance of extended states \cite{Extend-Stats}.

The main subject of the present work, is the effect of disorder correlations
on the transmission statistics in the asymptotic limit of large $L$\textbf{.
}Transmission coefficient of a long disordered system exhibits large
sample-to-sample fluctuations and is not a self-averaging quantity. General
properties of the transmission distribution have been considered in the past
using the composition rule for the transmission of a one-dimensional chain of
statistically identical and independent (or only weakly dependent) random
scatterers \cite{Anderson-80},\cite{Shapiro-88}. It was shown that a
convenient variable to deal with is $\ln T\left(  L\right)  $, since it can be
represented as a sum of independent (or weakly dependent) random variables,
for which the conditions of the Central Limit theorem, or its modification for
weakly correlated variables, are valid. Then, by additivity of cumulants of
independent variables, it follows that all cumulants of the $\ln
T$-distribution, $P\left(  \ln T;L\right)  $, grow at most linearly in the
system length $L$:%
\begin{equation}
\left\langle \left\langle \left[  \ln T^{-1/2}\left(  L\right)  \right]
^{n}\right\rangle \right\rangle =c_{n}L+o\left(  L\right)  ,\quad
n=1,2,\ldots. \label{Cn coefs-def_0}%
\end{equation}
Here $\left\langle \left\langle X^{n}\right\rangle \right\rangle $ denotes
$n$th-order cumulant of the quantity $X$, and the asymptotic cumulant
coefficients $c_{n}$ are constants, which depend on the microscopic properties
of the system. The above result applies to systems with both uncorrelated and
correlated random potentials, provided the considered system can be divided
into blocks, which are much longer than the disorder correlation scale and can
be treated as nearly independent scatterers.

Thus, the general form of the large-$L$ limit of the transmission distribution
is well understood, and the remaining questions are about the dependence of
the cumulant coefficients $c_{n}$ on the parameters of the problem. According
to Eq. (\ref{Cn coefs-def_0}), transmission statistics can be discussed at the
following three levels of "resolution". First, there exists a self-averaging
quantity $\left(  2L\right)  ^{-1}\ln T^{-1}$, which has a non-random limit%
\begin{equation}
\lim_{L\rightarrow\infty}\left(  2L\right)  ^{-1}\ln T^{-1}=c_{1}\equiv\gamma.
\end{equation}
At this level, all the information about the transmission fluctuations is lost
in the asymptotic limit $L\rightarrow\infty$. Next, one can define a variable
$\left(  \ln T-\left\langle \ln T\right\rangle \right)  /\left(
4c_{2}L\right)  ^{1/2}$, where $\left\langle \ln T\right\rangle =2c_{1}L$, for
which the Central Limit theorem holds, and whose cumulants of the order
$n\geq3$ vanish in the asymptotic limit $L\rightarrow\infty$ as $L^{-\left(
n-2\right)  /2}$. Therefore, this veriable has a limiting Gaussian
distribution with zero mean and unit variance. Limiting probability
distributions and their relation to universality and scaling have been
discussed in the past \cite{Shapiro-88},\cite{Shapiro-87}. In the present
context of a 1D problem, only two model-dependent parameters, $c_{1}$ and
$c_{2}$, are "remembered" in the limiting distribution of $\left(  \ln
T-\left\langle \ln T\right\rangle \right)  /\left(  4c_{2}L\right)  ^{1/2}$.
This fact is referred to as "two-parameter scaling" \cite{Shapiro-88}. In
special conditions, namely when the combined scatterers are weak and their
reflection phases are distributed uniformly, the composition rule yields
\thinspace$c_{1}=c_{2}$ up to weak disorder corrections, which is a
manifestation of single parameter scaling (SPS) \cite{Anderson-80}%
,\cite{Shapiro-88}.

Finally, for large but finite $L$, one can study the non-Gaussian corrections
to the limiting Gaussian distribution, which are characterized by the
asymptotic cumulant coefficients $c_{n}$, $n\geq3$. The latter are responsible
for the extreme fluctuations of the transmission coefficient. These large
deviations can also be studied in terms of the generalized LEs, which describe
asymptotic growth rates of the moments of the inverse transmission coefficient
$T^{-1}$\ and can be expressed as a sum over $c_{n}$'s \cite{Pikovsky-03}.

Under conditions of weak scattering and phase randomization, in addition to
the SPS relation \thinspace$c_{1}=c_{2}$, it was found in some models of
disorder that cumulant ratios $c_{n}/c_{1}$, $n\geq3$, vanish up to the weak
disorder corrections \cite{Schomerus-02},\cite{Petri-96-97}. Vanishing of
higher cumulants is also indicated by the coincidence of LE $\gamma$ and the
second order generalized LE $\Lambda$ in the lowest order of the weak disorder
expansion, which is valid for a wide variety of 1D disordered models
\cite{Pikovsky-03},\cite{LGP-Introduction},\cite{Tessiery-Israilev-01}%
,\cite{Tessieri-02}. Thus, quite generally, cumulant coefficients $c_{n}$
satisfy%
\begin{equation}
c_{2}/c_{1}=1,\quad c_{n}/c_{1}=0,\quad n\geq3, \label{SPS rel}%
\end{equation}
which holds up to the weak disorder corrections.

Considerable analytical and numerical effort was devoted to verification of
SPS and determining limits of its validity for various models of disorder
\cite{Schomerus-02},\cite{Shapiro-88},\cite{Petri-96-97},\cite{Abrikosov-81}%
,\cite{Altshuler-01},\cite{Tamura-93},\cite{Zaslavsky-97}. Usually, SPS holds
for sufficiently weak disorder, when the localization length is much larger
than other scales in the problem (see also discussion in Ref.
\cite{Altshuler-01}).

Deviations from relations (\ref{SPS rel}) were studied for uncorrelated
disorder \cite{Pikovsky-03},\cite{Schomerus-02},\cite{Schomerus-03} as well as
for some specific models of correlated disorder potentials \cite{Titov-05}%
,\cite{Deych-03}. In particular, for the continuous white noise model, it was
shown \cite{Schomerus-02} that $c_{2}/c_{1}=1$ for large positive energy
$\epsilon$ (weak disorder), becomes slightly different from unity near
$\epsilon=0$ and vanishes for large negative $\epsilon$, while the
"non-Gaussian" ratios $c_{n\geq3}/c_{1}$ have numerical values of about a
fraction of unity near $\epsilon=0$, and vanish for large negative or positive
$\epsilon$ (energy scale is determined by strength of disorder). Thus, in the
white noise model, unless $\epsilon$ is sufficiently large and negative, the
relative width of the Gaussian bulk of the $\ln T$-distribution depends weakly
on the disorder strength, while finite higher cumulants indicate that some
non-Gaussian corrections to the distribution tails develop in strong disorder.
The latter was also observed in Ref. \cite{Pikovsky-03}, where distribution of
local Lyapunov exponents was studied in terms of the generalized LEs for a
continuous $\delta$-correlated disorder.

Presence of disorder correlations can significantly enhance deviation from
relations (\ref{SPS rel}). For instance, one finds $\left\vert c_{2}%
/c_{1}-1\right\vert \sim1$ near certain energies of the tight-binding model
with a weak short-range correlated disorder \cite{Titov-05}. Another example
of the effect of correlations is scaling relation $c_{2}/c_{1}\sim c_{1}R_{c}%
$, obtained in Ref. \cite{Deych-03} for strong exponentially correlated
dichotomous disorder with a correlation scale $R_{c}$. This scaling, which
replaces Eq. (\ref{SPS rel}), is valid when $c_{1}R_{c}\gg1$, which means
$c_{2}/c_{1}\gg1$, so that Gaussian bulk of the distribution $P\left(  \ln
T;L\right)  $ becomes much broader than in weak or white noise disorder.

These special examples suggest that disorder correlations can have a
significant effect on the transmission statistics. One can distinguish two
limiting regimes. In weak disorder, which can be treated perturbatively, LE
depends strongly on the correlation scale $R_{c}$ \cite{LGP-Introduction}%
,\cite{Israilev-99},\cite{Gurevich},\cite{Palencia-09}. However, it usually
remains the only parameter of the transmission distribution, and relations
(\ref{SPS rel}) stay valid \cite{Petri-96-97},\cite{Tamura-93}%
,\cite{Zaslavsky-97}. Another special regime, which can be called a
semiclassical regime of strong disorder, occurs when typical random potential
barriers are higher than energy $\epsilon$ and sufficiently broad, so that
tunneling probability already through a single random barrier becomes small.
In this case localization is dominated by the under-barrier tunneling, and one
can generalize the semiclassical approach of Ref. \cite{Deych-03} to study the
distribution of $\ln T\left(  L\right)  $ in terms of the statistics of
disorder excursions above the level $V\left(  x\right)  =\epsilon$. In the
present work we use this approximation for the strong disorder regime to show
that $c_{n}/c_{1}\sim\left(  c_{1}R_{c}\right)  ^{n-1}$.

The main focus of the present research is, however,\ on the effect of disorder
correlations on the transmission distribution at the transition between these
two limiting regimes. We consider model (\ref{Schrodinger}) with Gaussian
disorder $V\left(  x\right)  $, whose correlation function is arbitrary,
except for the assumption that it is parametrized with a finite correlation
scale $R_{c}$. The transmission statistics is studied in terms of the
dimensionless ratios $c_{2}/c_{1}$, $c_{3}/c_{1}$ and $\Lambda/c_{1}$.
Quantity $\Lambda/c_{1}$ characterizes extreme fluctuations of the inverse
transmission coefficient $T^{-1}$, since it is related to the ratio of the
average to the typical values of $T^{-1}$. In addition, it can be used to show
violation of relations (\ref{SPS rel}), because $\Lambda/c_{1}=1$ when these
relations hold \cite{Pikovsky-03},\cite{Tessieri-02}.

To calculate $\Lambda$, which can not be treated exactly in a general case of
correlated disorder, we develop a small-$R_{c}$ approximation \cite{Klatzkin}.
A simple regularization is proposed to extend this method to the power law
correlations, for which a standard approximation turns out to be inapplicable.
Numerical simulations are used to verify the analytical approximation for
$\Lambda$ and to calculate cumulant coefficients $c_{1,2,3}$ for different
types of correlations and for different values of $R_{c}$ (and $\epsilon>0$).
As expected, the obtained results for $c_{2}/c_{1}$, $c_{3}/c_{1}$ and
$\Lambda/c_{1}$ show that SPS and Eq. (\ref{SPS rel}) hold in the weak
disorder limit. Beyond this regime, we find that disorder correlations
strongly enhance deviations from relations (\ref{SPS rel}), as compared to the
case of white noise, and lead to increase of both typical and extreme
fluctuations of transmission. Dependence of these effects on the parameters
$kR_{c}$ and $c_{1}R_{c}$ is discussed ($k=\sqrt{\epsilon}$ is the
wavenumber). In general, effect of disorder correlations on the transmission
statistics becomes important when $c_{1}R_{c}$, i.e. the ratio between the
correlation and the localization lengths, is comparable to or greater than unity.

The outline of the paper is as follows. In Sec. \ref{Sect: Model} we define
the model in question and discuss in some more detail the standard and the
generalized Lyapunov exponents. The small-$R_{c}$ approximation for the
generalized LE $\Lambda$ is developed in Sec. \ref{Sect: GLE}. First, using
the analogy of Eq. (\ref{Schrodinger}) with the Langevin equation, equations
for the second moments of $\psi\left(  x\right)  $ are obtained by means of
the Furutsu-Novikov formula. This leads to an infinite hierarchy of the
integro-differential equations for the second moment of the wave function and
its functional derivatives \cite{Hanggi-95},\cite{Klatzkin}. Then, the
small-$R_{c}$ closure approximation is applied and the result is compared to
numerical simulations. In Sec. \ref{Sect: Semiclass} we obtain scaling
relations for $c_{n}/c_{1}$ in the semiclassical regime of strong disorder. In
Sec. \ref{Sect: Numerical results} analytical and numerical results for
$\Lambda$ and $c_{n}$ are used to discuss properties of the asymptotic
distribution of the transmission coefficient and deviation from relations
(\ref{SPS rel}). The results are summarized in Sec. \ref{Sect: Coclus}. Known
results on the generalized LE for the $\delta$-correlated disorder and on the
Born approximation are presented in appendices \ref{Sect: GLE for WN} and
\ref{Sect: Born Approx} respectively. Method of numerical simulation is
explained in Appendix \ref{Sect: Numrics}. Some auxiliary formulae for the
semiclassical treatment of strong disorder are presented in Appendix
\ref{Sect: Auxiliary}.

\section{Basic equations\label{Sect: Model}}

We consider one-dimensional Schr\"{o}dinger equation%
\begin{equation}
-\frac{d^{2}\psi}{dx^{2}}+V(x)\psi=\epsilon\psi, \label{eqn1}%
\end{equation}
where $\epsilon$ is energy and $V(x)$ is zero-mean Gaussian disorder
potential. The latter is completely described by its two-point correlation
function $C_{2}\left(  x-x^{\prime}\right)  \equiv\left\langle V(x)V(x^{\prime
})\right\rangle $, written as%
\begin{equation}
C_{2}\left(  x-x^{\prime}\right)  =V_{0}^{2}\Gamma\left(  \frac{x-x^{\prime}%
}{R_{c}}\right)  , \label{eq. DO properties}%
\end{equation}
where $V_{0}^{2}$ and $R_{c}$ are disorder variance and correlation length,
and the dimensionless function $\Gamma(x)$ decays on the scale of unity and is
chosen so that%
\begin{equation}
\Gamma(0)=1,\quad\int_{0}^{+\infty}\Gamma(x)dx=1.
\label{normalization of Gamma2}%
\end{equation}
The white noise model is obtained by taking the limit $R_{c}\rightarrow0$,
while keeping $R_{c}V_{0}^{2}$ constant:%
\begin{equation}
\lim_{R_{c}\longrightarrow0}C_{2}\left(  x-x^{\prime}\right)  =g\delta
(x-x^{\prime}), \label{eqn4}%
\end{equation}
where we have introduced disorder intensity parameter%
\begin{equation}
g\equiv\int_{-\infty}^{+\infty}C_{2}(x)dx=2R_{c}V_{0}^{2}.
\label{eq. DO parameter}%
\end{equation}
The transmission properties of a one-dimensional system of finite length $L$
are described by transfer matrix \cite{Mello-Kumar}
\begin{equation}
\frac{1}{\sqrt{T}}\left[
\begin{array}
[c]{cc}%
e^{i\phi_{t}} & -\sqrt{1-T}e^{-i\left(  \phi_{r}-\phi_{t}\right)  }\\
-\sqrt{1-T}e^{i\left(  \phi_{r}-\phi_{t}\right)  } & e^{-i\phi_{t}}%
\end{array}
\right]  , \label{Transfer Matr}%
\end{equation}
which relates amplitudes of the incident and the outgoing waves (a time
reversal invariance is implied). Here $\phi_{r}$ and $\phi_{t}$ are reflection
and transmission phases respectively. In the following sections, generalized
LE $\Lambda$ and asymptotic cumulant coefficients $c_{n}$ are studied in terms
of the amplitude $A^{2}\left(  x\right)  =\left\vert \psi\left(  x\right)
\right\vert ^{2}+\left\vert \psi^{\prime}\left(  x\right)  \right\vert ^{2}$,
where $\psi\left(  x\right)  $ is a solution of (\ref{eqn1}) satisfying some
generic initial conditions, e.g. $\psi\left(  0\right)  =1$, $\psi^{\prime
}\left(  0\right)  =0$. The obtained results are valid also for the
transmission coefficient $T$, since in the asymptotic limit $L\rightarrow
\infty$ it becomes statistically equivalent to the amplitude $A\left(
x\right)  $. This is because $A\left(  x\right)  $ is expressed in terms of
the transfer matrix (\ref{Transfer Matr}), which asymptotically factorizes
into the product of large factor $T^{-1}$ and statistically independent of it
matrix of the order of unity. Such factorization takes place, since for $L$
much larger than the localization length, phases $\phi_{r}$ and $\phi_{t}$
become statistically (almost \cite{Note-T-phase-Coorel}) independent of $T$,
while coefficient $T$ becomes exponentially small in most realizations of
disorder. Thus, for large $L$, one can write%
\begin{equation}
\ln A^{2}\left(  L\right)  =\ln T^{-1}\left(  L\right)  +O\left(
L^{0}\right)  ,
\end{equation}
where the term $O\left(  L^{0}\right)  $, which absorbs phases $\phi_{r,t}$
and the initial conditions for $\psi\left(  x\right)  $, is statistically
independent of the first one, and its cumulants are of the order of unity.
Therefore, using additivity of cumulants of independent variables, one has
(cf. Eq.(\ref{Cn coefs-def_0}))%
\begin{equation}
c_{n}\equiv\lim_{L\rightarrow\infty}\frac{\left\langle \left\langle \ln
^{n}T^{-1}\right\rangle \right\rangle }{2^{n}L}=\lim_{L\rightarrow\infty}%
\frac{\left\langle \left\langle \ln^{n}A^{2}\right\rangle \right\rangle
}{2^{n}L}. \label{Cn coefs-def}%
\end{equation}
Thus, for large $L$, inverse transmission coefficient $T^{-1}$ is
statistically equivalent to the amplitude $A^{2}\left(  x=L\right)  $, and,
thus, to the local LE $\tilde{\gamma}\left(  x=L\right)  $, defined in the
introduction. In particular, LE $\gamma$ coincides with $c_{1}$, and in the
following we will use the latter notation.

As already mentioned in the introduction, cumulant coefficients $c_{n}$
describe asymptotic form of the distribution $P\left(  \ln T;L\right)  $.
Introducing the dimensionless length $l\equiv c_{1}L$ measured in units of the
localization length $c_{1}^{-1}$, cumulants of $\ln T$ can be rewritten as
(cf. Eq. (\ref{Cn coefs-def_0}))%
\begin{equation}
\left\langle \left\langle \ln^{n}T^{-1/2}\left(  l\right)  \right\rangle
\right\rangle =\frac{c_{n}}{c_{1}}l+O\left(  l^{0}\right)  .
\label{T_cums_dimless}%
\end{equation}
Therefore, it is convenient to discuss transmission distributions in terms of
the dimensionless ratios $c_{n}/c_{1}$, since meaningful comparison of
transmission fluctuations in different models of disorder should be done for
the same dimensionless system length $l$. Then, for instance, relative
fluctuation of $\ln T$ is expressed as $\left\langle \left\langle \ln
^{2}T\right\rangle \right\rangle /\left\langle \ln T\right\rangle ^{2}=\left(
c_{2}/c_{1}\right)  /l$.

Another quantity which characterizes transmission distribution is second order
generalized LE $\Lambda$ defined by \cite{Paladin-87}%
\begin{equation}
\Lambda=\underset{L\rightarrow\infty}{\lim}\frac{1}{4L}\ln\left\langle
A^{2}\left(  L\right)  \right\rangle =\underset{L\rightarrow\infty}{\lim}%
\frac{1}{4L}\ln\left\langle T^{-1}\left(  L\right)  \right\rangle ,
\label{eq. generalized LE}%
\end{equation}
Since $\ln\left\langle A^{q}\left(  x\right)  \right\rangle =\ln\left\langle
e^{q\ln A\left(  x\right)  }\right\rangle $ is a cumulant generating function
of the distribution $P\left(  \ln A;x\right)  $, the generalized LE is related
to all cumulant coefficients $c_{n}$ by the expression%
\begin{equation}
\Lambda=\frac{1}{4}\sum_{n=1}^{\infty}\frac{2^{n}}{n!}c_{n}. \label{GLE2Cums}%
\end{equation}
For large $L$, value of $\left\langle T^{-1}\left(  L\right)  \right\rangle $
is dominated by the long tail of the $T^{-1}$-distribution, which becomes
highly skewed and heavy-tailed. Therefore, rather than representing typical
values of $T^{-1}$, generalized LE $\Lambda$ provides useful complementary
information about the low-$T$ tail of its distribution. To this end, it is
convenient to define the following quantity
\begin{equation}
\rho\equiv\frac{\Lambda}{c_{1}}=\frac{1}{4}\sum_{n=1}^{\infty}\frac{2^{n}}%
{n!}\frac{c_{n}}{c_{1}}, \label{Ro_def}%
\end{equation}
whose meaning is twofold. On the one hand, deviation of the value of $\rho$
from unity gives an integral measure of deviation from the universal weak
disorder relations (\ref{SPS rel}), since $\rho=1$ when these relations hold
(cf. Ref. \cite{Pikovsky-03}). On the other hand, $\rho$ can be used to
characterizes the extreme relative fluctuations of the inverse transmission
coefficient $T^{-1}$ in terms of the ratio between the mean and the typical
values of $T^{-1}$:
\begin{equation}
\frac{\left\langle T^{-1}\right\rangle }{\exp\left\langle \ln T^{-1}%
\right\rangle }=e^{\left(  2\rho-1\right)  2l}. \label{Extreme_fluct_measure}%
\end{equation}

\section{Generalized Lyapunov Exponent $\Lambda$\label{Sect: GLE}}

\subsection{Equation of motion for moments\label{Sect: Langevin}}

The linearity of the Schr\"{o}dinger equation (\ref{eqn1}) allows one to
obtain closed-form equations for the $2n$-order products $\left[
\psi(x)\right]  ^{k}\left[  \psi^{\prime}\left(  x\right)  \right]  ^{l}\,$,
where$~~k+l=2n$,~~$k,l=0,1,2,\dots\,$. To this end, we rewrite Eq.
(\ref{eqn1}) in the form of the Langevin equation. The $x$ coordinate is
considered as a formal time on the half axis $x\equiv t\in\lbrack0,\infty)$
and the dynamical variables $u(t)\equiv\psi\left(  x\right)  ,~v(t)\equiv
\psi^{\prime}\left(  x\right)  $ are introduced. In these variables the
Langevin equation reads
\begin{equation}
\partial_{t}u=v\,,~~\partial_{t}v=[V(t)-\epsilon]u\,, \label{Evolut for u,v}%
\end{equation}
where $V(t)$ is now the correlated noise. We need to consider an initial value
problem for the second order moments $\left\langle \left\vert u\right\vert
^{2}\right\rangle $, $\sqrt{2^{-1}}\left\langle u^{\ast}v+uv^{\ast
}\right\rangle $, $\left\langle \left\vert v\right\vert ^{2}\right\rangle $,
whose asymptotic exponential growth rate gives the generalized LE $\Lambda$,
Eq. (\ref{eq. generalized LE}). Introducing vector%
\begin{equation}
\mathbf{Y}=(\left\vert u\right\vert ^{2},\sqrt{2^{-1}}\left[  u^{\ast
}v+uv^{\ast}\right]  ,\left\vert v\right\vert ^{2})^{T}, \label{Y_def}%
\end{equation}
and using Eq. (\ref{Evolut for u,v}), one obtains%
\begin{equation}
\partial_{t}\mathbf{Y}=\left(  \mathcal{C}+V\left(  t\right)  \mathcal{D}%
\right)  \mathbf{Y},\ \mathbf{Y}\left(  0\right)  =\mathbf{Y}_{0},
\label{Evolut for moments}%
\end{equation}
where%
\begin{equation}
\mathcal{C}=\sqrt{2}%
\begin{bmatrix}
0 & 1 & 0\\
-\epsilon & 0 & 1\\
0 & -\epsilon & 0
\end{bmatrix}
;\ \mathcal{D}=\sqrt{2}%
\begin{bmatrix}
0 & 0 & 0\\
1 & 0 & 0\\
0 & 1 & 0
\end{bmatrix}
\label{C and D}%
\end{equation}
and $\mathbf{Y}_{0}$ is an initial condition. To obtain equation for
$\left\langle \mathbf{Y}\left(  t\right)  \right\rangle $, Eq.
(\ref{Evolut for moments}) is averaged over the disorder, which introduces
correlator $\left\langle V\left(  t\right)  \mathbf{Y}\left(  t\right)
\right\rangle $:
\begin{equation}
\partial_{t}\left\langle \mathbf{Y}\right\rangle =\mathcal{C}\left\langle
\mathbf{Y}\right\rangle +\mathcal{D}\left\langle V\left(  t\right)
\mathbf{Y}\right\rangle ,\quad\left\langle \mathbf{Y}\left(  0\right)
\right\rangle =\mathbf{Y}_{0}. \label{Evolut for <Y>}%
\end{equation}
Applying the Furutsu-Novikov formula \cite{Hanggi-95},\cite{Klatzkin} yields%
\begin{equation}
\partial_{t}\left\langle \mathbf{Y}\right\rangle =\mathcal{C}\left\langle
\mathbf{Y}\right\rangle +\int_{0}^{t}C_{2}\left(  t-\tau\right)
\mathcal{D}\left\langle \frac{\delta\mathbf{Y}\left(  t\right)  }{\delta
V\left(  \tau\right)  }\right\rangle d\tau, \label{Evolut for <Y> with N-F}%
\end{equation}
where $\frac{\delta\mathbf{Y}\left(  t\right)  }{\delta V\left(  \tau\right)
}$ is a functional derivative of $\mathbf{Y}\left(  t\right)  $ with respect
to $V\left(  \tau\right)  $ and $C_{2}\left(  t\right)  $ is the disorder
correlation function (\ref{eq. DO properties}). Since equation
(\ref{Evolut for <Y> with N-F}) contains a new quantity $\left\langle
\frac{\delta\mathbf{Y}\left(  t\right)  }{\delta V\left(  \tau\right)
}\right\rangle $, it is not closed with respect to $\left\langle
\mathbf{Y}\right\rangle $. An important exception is the case of the
uncorrelated disorder, which is considered in Appendix \ref{Sect: GLE for WN},
and will be used in the forthcoming analysis. In general case, to proceed
further, an additional equation for $\frac{\delta\mathbf{Y}\left(  t\right)
}{\delta V\left(  \tau\right)  }$ is required, which is obtained by
differentiating Eq. (\ref{Evolut for moments}) functionally with respect to
$V\left(  \tau\right)  $, and using the differentiation property $\frac
{\delta}{\delta V\left(  \tau\right)  }\frac{\partial}{\partial t}%
\mathbf{Y}\left(  t\right)  =\frac{\partial}{\partial t}\frac{\delta
\mathbf{Y}\left(  t\right)  }{\delta V\left(  \tau\right)  }$ \cite{Klatzkin}.
This yields
\begin{align}
\frac{\partial}{\partial t}  &  \frac{\delta\mathbf{Y}\left(  t\right)
}{\delta V\left(  \tau\right)  }=\mathcal{D}\mathbf{Y}\left(  \tau\right)
\delta\left(  t-\tau\right)  +\nonumber\\
&  +\Theta\left(  t-\tau\right)  \left[  \mathcal{C}+V\left(  t\right)
\mathcal{D}\right]  \frac{\delta\mathbf{Y}\left(  t\right)  }{\delta V\left(
\tau\right)  },\quad\frac{\delta\mathbf{Y}\left(  t<\tau\right)  }{\delta
V\left(  \tau\right)  }=0, \label{Evolut for dY/dh (0)}%
\end{align}
where $\Theta\left(  t-\tau\right)  $ is the Heaviside step function. For
$t>\tau$, Eq. (\ref{Evolut for dY/dh (0)}) can be rewritten as%
\begin{equation}
\frac{\partial}{\partial t}\frac{\delta\mathbf{Y}\left(  t\right)  }{\delta
V\left(  \tau\right)  }=\left[  \mathcal{C}+V\left(  t\right)  \mathcal{D}%
\right]  \frac{\delta\mathbf{Y}\left(  t\right)  }{\delta V\left(
\tau\right)  },\ \frac{\delta\mathbf{Y}\left(  \tau^{+}\right)  }{\delta
V\left(  \tau\right)  }=\mathcal{D}\mathbf{Y}\left(  \tau\right)  .
\label{Evolut for dY/dh}%
\end{equation}
The causality property $\frac{\delta\mathbf{Y}\left(  t\right)  }{\delta
V\left(  \tau>t\right)  }=0$ was used in derivation of Eq.
(\ref{Evolut for dY/dh (0)}). It follows from the fact that $\mathbf{Y}\left(
t\right)  $ obeys first order differential equation (\ref{Evolut for moments})
with an initial condition at $t=0$ and, thus, is independent of $V\left(
\tau\right)  $ at the time $\tau$ later than $t$. Then, integration of the
differential equation (\ref{Evolut for dY/dh (0)}) over $t$ from some
$t_{1}<\tau$ up to $\tau^{+}$ yields the initial condition in Eq.
(\ref{Evolut for dY/dh}).

Note that functional derivative $\frac{\delta\mathbf{Y}\left(  t\right)
}{\delta V\left(  \tau\right)  }$ in Eq. (\ref{Evolut for dY/dh}) and vector
$\mathbf{Y}\left(  t\right)  $ in Eq. (\ref{Evolut for moments}) obey the same
differential equation, but with different initial condition. Due to this
similarity, moments $\left\langle \frac{\delta\mathbf{Y}\left(  t\right)
}{\delta V\left(  \tau\right)  }\right\rangle $ and $\left\langle
\mathbf{Y}\left(  t\right)  \right\rangle $ grow with the same asymptotic rate
for large $\left(  t-\tau\right)  $ and $t$ respectively. The difference
between the two quantities is that the initial condition for $\frac
{\delta\mathbf{Y}\left(  t\right)  }{\delta V\left(  \tau\right)  }$ depends
on $\mathbf{Y}\left(  \tau\right)  $ and, therefore, it is correlated to
$V\left(  t\right)  $.

Averaging Eq. (\ref{Evolut for dY/dh}) over the disorder and using the
Furutsu-Novikov formula, one obtains the following equation for the moment
$\left\langle \frac{\delta\mathbf{Y}\left(  t\right)  }{\delta V\left(
\tau\right)  }\right\rangle $:%
\begin{align}
\frac{\partial}{\partial t}\left\langle \frac{\delta\mathbf{Y}\left(
t\right)  }{\delta V\left(  \tau\right)  }\right\rangle  &  =\mathcal{C}%
\left\langle \frac{\delta\mathbf{Y}\left(  t\right)  }{\delta V\left(
\tau\right)  }\right\rangle +\nonumber\\
+  &  \int_{0}^{t}C_{2}\left(  t-\tau^{\prime}\right)  \mathcal{D}\left\langle
\frac{\delta^{2}\mathbf{Y}\left(  t\right)  }{\delta V\left(  \tau^{\prime
}\right)  \delta V\left(  \tau\right)  }\right\rangle d\tau^{\prime}
\label{Evolut for <dY/dh> with N-F}%
\end{align}
with the initial condition
\begin{equation}
\left\langle \frac{\delta\mathbf{Y}\left(  t=\tau^{+}\right)  }{\delta
V\left(  \tau\right)  }\right\rangle =\mathcal{D}\left\langle \mathbf{Y}%
\left(  \tau\right)  \right\rangle .
\label{I.C. for Evolut for <dY/dh> with N-F}%
\end{equation}
Thus, Eq. (\ref{Evolut for <dY/dh> with N-F})\ is again not closed with
respect to $\left\langle \frac{\delta\mathbf{Y}\left(  t\right)  }{\delta
V\left(  \tau\right)  }\right\rangle $, because it contains an additional
quantity $\left\langle \frac{\delta^{2}\mathbf{Y}\left(  t\right)  }{\delta
V\left(  \tau^{\prime}\right)  \delta V\left(  \tau\right)  }\right\rangle $.
Then, the above procedure can be iterated repeatedly to obtain an infinite
hierarchy of coupled differential equations involving moments of higher
functional derivatives of $\mathbf{Y}$ with respect to $V\left(  t\right)  $
\cite{Hanggi-95},\cite{Klatzkin}.

\subsection{Small-$R_{c}$ decoupling approximation}

To make progress with the infinite hierarchy of coupled differential equation,
the first two of which are equations (\ref{Evolut for <Y> with N-F}) and
(\ref{Evolut for <dY/dh> with N-F}), one usually resorts to some closure
approximation \cite{Hanggi-95},\cite{Klatzkin}. In this work we apply a
small-$R_{c}$ approximation to close the second equation
(\ref{Evolut for <dY/dh> with N-F}) with respect to $\left\langle \frac
{\delta\mathbf{Y}\left(  t\right)  }{\delta V\left(  \tau\right)
}\right\rangle $. This allows to express $\left\langle \frac{\delta
\mathbf{Y}\left(  t\right)  }{\delta V\left(  \tau\right)  }\right\rangle $ in
terms of the initial condition $\mathcal{D}\left\langle \mathbf{Y}\left(
\tau\right)  \right\rangle $. Then, substituting it into Eq.
(\ref{Evolut for <Y> with N-F}), one obtains a closed integro-differential
equation for $\left\langle \mathbf{Y}\right\rangle $.

A formal solution for $\frac{\delta\mathbf{Y}\left(  t\right)  }{\delta
V\left(  \tau\right)  }$ is obtained from Eq. (\ref{Evolut for dY/dh}) in the
form%
\begin{equation}
\frac{\delta\mathbf{Y}\left(  t\right)  }{\delta V\left(  \tau\right)
}=\mathcal{K}\left(  t,\tau\right)  \mathcal{D}\mathbf{Y}\left(  \tau\right)
, \label{Sol dY/dh with K}%
\end{equation}
where the evolution operator%
\begin{equation}
\mathcal{K}\left(  t,\tau\right)  \equiv\mathbf{T\exp}\int_{\tau}^{t}%
d\tau\left(  \mathcal{C}+V\left(  \tau\right)  \mathcal{D}\right)
\end{equation}
is the $T$-ordered exponent. Disorder average of Eq. (\ref{Sol dY/dh with K})
introduces a correlator $\left\langle \mathcal{K}\left(  t,\tau\right)
\mathcal{D}\mathbf{Y}\left(  \tau\right)  \right\rangle $. Application of the
decoupling, or mean field, approximation yields%
\begin{equation}
\left\langle \frac{\delta\mathbf{Y}\left(  t\right)  }{\delta V\left(
\tau\right)  }\right\rangle \approx\left\langle \mathcal{K}\left(
t-\tau\right)  \right\rangle \mathcal{D}\left\langle \mathbf{Y}\left(
\tau\right)  \right\rangle , \label{Decoupling}%
\end{equation}
where we can write $\left\langle \mathcal{K}\left(  t,\tau\right)
\right\rangle =\left\langle \mathcal{K}\left(  t-\tau\right)  \right\rangle $
due to the stationarity of $V\left(  t\right)  $. Terms, neglected in Eq.
(\ref{Decoupling}) are given formally by the generalization of the
Furutsu-Novikov formula for the correlation between two functionals of the
random Gaussian process \cite{Hanggi-95}:%
\begin{gather}
\left\langle F\left\{  V\right\}  G\left\{  V\right\}  \right\rangle
=\left\langle F\left\{  V\right\}  \right\rangle \left\langle G\left\{
V\right\}  \right\rangle +\quad\quad\quad\quad\quad\quad\quad\nonumber\\
+\sum_{n=1}^{\infty}\frac{1}{n!}\idotsint\left\langle \frac{\delta
^{n}F\left\{  V\right\}  }{\delta V\left(  t_{n}\right)  \cdots\delta V\left(
t_{1}\right)  }\right\rangle \times\quad\quad\quad\nonumber\\
\times\left\langle \frac{\delta^{n}G\left\{  V\right\}  }{\delta V\left(
s_{n}\right)  \cdots\delta V\left(  s_{1}\right)  }\right\rangle
{\displaystyle\prod_{i=1}^{n}}
C_{2}\left(  t_{i}-s_{i}\right)  dt_{i}ds_{i}. \label{N-F for functionals}%
\end{gather}
In general case, when functionals $F\left\{  V\right\}  $ and $G\left\{
V\right\}  $ depend on the noise at simultaneous times, the decoupling
approximation does not rely on small correlation radius, but depends on other
parameters of the problem, e.g., small noise intensity \cite{Hanggi-95}. The
present case is special, because $\mathbf{Y}\left(  \tau\right)  $ and
$\mathcal{K}\left(  t,\tau\right)  $ depend on $V\left(  t^{\prime}\right)  $
at separate time intervals, $0<t^{\prime}<\tau$ and $\tau<t^{\prime}<t$
respectively. Therefore, the decoupling approximation for $\left\langle
\mathcal{K}\left(  t,\tau\right)  \mathcal{D}\mathbf{Y}\left(  \tau\right)
\right\rangle $ is also justified for sufficiently small $R_{c}$. In
particular, equation (\ref{Decoupling}) becomes exact in the white noise limit
(cf. Eq. (\ref{K for WN}) below). A small-$R_{c}$ perturbative expansion of
the\ $n=1$ term in Eq. (\ref{N-F for functionals}) shows that it can be
neglected if both $kR_{c}$ and $\Lambda R_{c}$ are small compared to unity.

Substitution of the decoupling approximation (\ref{Decoupling}) into Eq.
(\ref{Evolut for <Y> with N-F}) yields%
\begin{align}
\frac{\partial}{\partial t}\left\langle \mathbf{Y}\left(  t\right)
\right\rangle  &  =\mathcal{C}\left\langle \mathbf{Y}\left(  t\right)
\right\rangle +\nonumber\\
+  &  \mathcal{D}\int_{0}^{t}C_{2}\left(  t-\tau\right)  \left\langle
\mathcal{K}\left(  t-\tau\right)  \right\rangle \mathcal{D}\left\langle
\mathbf{Y}\left(  \tau\right)  \right\rangle d\tau,
\label{Evolut for <Y> with decoupling}%
\end{align}
Note that inserting $\left\langle \mathbf{Y}\left(  t\right)  \right\rangle
=\left\langle \mathcal{K}\left(  t\right)  \right\rangle \mathbf{Y}_{0}$ into
Eq. (\ref{Evolut for <Y> with decoupling}), one arrives at closed nonlinear
equation for $\left\langle \mathcal{K}\left(  t\right)  \right\rangle $.
Alternatively, assuming in Eq. (\ref{Evolut for <Y> with decoupling}) some
explicit approximation for $\left\langle \mathcal{K}\left(  t-\tau\right)
\right\rangle $, which will be defined later, one obtains closed linear
equation for $\left\langle \mathbf{Y}\left(  t\right)  \right\rangle $. It can
be solved in a standard way by the Laplace transform, while its large time
($t\gg R_{c}$) eigenvalues can be determined by substituting the asymptotic
solution in the form $\left\langle \mathbf{Y}\left(  t\right)  \right\rangle
=\mathbf{Y}_{\infty}e^{4\tilde{\Lambda}t}$, where $\mathbf{Y}_{\infty}$ is a
constant vector and $\tilde{\Lambda}$ is an eigenvalue to be found. For $t\gg
R_{c}$, owing to the decay of the correlation function $C_{2}\left(
t-\tau\right)  $, the lower limit of integration in Eq.
(\ref{Evolut for <Y> with decoupling}) can be replaced with $-\infty$. Then,
changing the integration variable as $\tau\rightarrow\left(  t-\tau\right)  $,
one obtains a stationary eigenvalue problem, from which the generalized LE
$\Lambda$ is found as the largest real root of the characteristic equation%
\begin{equation}
\det\left[  \mathcal{C}+\mathcal{D}\left(  \int_{0}^{\infty}C_{2}\left(
\tau\right)  \left\langle \mathcal{K}\left(  \tau\right)  \right\rangle
e^{-4\tilde{\Lambda}\tau}d\tau\right)  \mathcal{D}-4\tilde{\Lambda}\right]
=0. \label{EigVal eq for Approxs}%
\end{equation}
Equation (\ref{EigVal eq for Approxs}) makes sense only when $g\equiv2\int
_{0}^{+\infty}C_{2}(x)dx$ is finite, which is satisfied by assumption
(\ref{normalization of Gamma2}). This condition is important, since the
correct asymptotic growth of $\left\langle \mathcal{K}\left(  \tau\right)
\right\rangle $ is proportional to $e^{4\Lambda\tau}$ (see note after Eq.
(\ref{Evolut for dY/dh})). Therefore, for large $\tau$, it should be exactly
canceled by the factor $e^{-4\tilde{\Lambda}\tau}$ in Eq.
(\ref{EigVal eq for Approxs}), when $\tilde{\Lambda}$ is equal to $\Lambda$.\ 

\subsection{White noise approximation for $\left\langle \frac{\delta
\mathbf{Y}\left(  t\right)  }{\delta V\left(  \tau\right)  }\right\rangle
$\label{Sect: WN approx}}

For $\left\langle \mathcal{K}\left(  \tau\right)  \right\rangle $ in Eq.
(\ref{EigVal eq for Approxs}) we apply a standard white noise approximation
\cite{Klatzkin} by replacing the true correlation function $C_{2}\left(
t-\tau^{\prime}\right)  $ in equation (\ref{Evolut for <dY/dh> with N-F}) for
$\left\langle \frac{\delta\mathbf{Y}\left(  t\right)  }{\delta V\left(
\tau\right)  }\right\rangle $ with its white noise limit $g\delta\left(
t-\tau^{\prime}\right)  $, where $g=2R_{c}V_{0}^{2}$, cf. Eq. (\ref{eqn4}).
The integral in Eq. (\ref{Evolut for <dY/dh> with N-F}) becomes%
\begin{align}
\int_{0}^{t}C_{2}\left(  t-\tau^{\prime}\right)   &  \mathcal{D}\left\langle
\frac{\delta^{2}\mathbf{Y}\left(  t\right)  }{\delta V\left(  \tau^{\prime
}\right)  \delta V\left(  \tau\right)  }\right\rangle d\tau^{\prime}%
\approx\nonumber\\
&  \mathcal{D}^{2}\left\langle \frac{\delta\mathbf{Y}\left(  t\right)
}{\delta V\left(  \tau\right)  }\right\rangle \int_{0}^{t}C_{2}\left(
t-\tau^{\prime}\right)  d\tau^{\prime}, \label{WN approx}%
\end{align}
where the initial condition $\frac{\delta^{2}\mathbf{Y}\left(  t^{+}\right)
}{\delta V\left(  t\right)  \delta V\left(  \tau\right)  }=\mathcal{D}%
\frac{\delta\mathbf{Y}\left(  t\right)  }{\delta V\left(  \tau\right)  }$\ was
used (cf. Eq. (\ref{Evolut for dY/dh})). Approximation (\ref{WN approx}) is
justified when $\left\langle \frac{\delta^{2}\mathbf{Y}\left(  t\right)
}{\delta V\left(  \tau^{\prime}\right)  \delta V\left(  \tau\right)
}\right\rangle $ changes slowly on the scale of $R_{c}$. Since relevant
characteristic scales of the problem are wave number $k=\sqrt{\left\vert
\epsilon\right\vert }$ and the generalized LE $\Lambda$, the conditions to be
fulfilled are%
\begin{equation}
kR_{c}\ll1\ \text{and}\ \Lambda R_{c}\ll1, \label{validity conds}%
\end{equation}
i.e. the same conditions as for the decoupling approximation (\ref{Decoupling}%
). Inserting the white noise approximation (\ref{WN approx}) into Eq.
(\ref{Evolut for <dY/dh> with N-F}) yields the following closed equation for
$\left\langle \frac{\delta\mathbf{Y}\left(  t\right)  }{\delta V\left(
\tau\right)  }\right\rangle $:
\begin{align}
\frac{\partial}{\partial t}\left\langle \frac{\delta\mathbf{Y}\left(
t\right)  }{\delta V\left(  \tau\right)  }\right\rangle  &  =\left[
\mathcal{C}+\frac{g}{2}\mathcal{D}^{2}\right]  \left\langle \frac
{\delta\mathbf{Y}\left(  t\right)  }{\delta V\left(  \tau\right)
}\right\rangle ,\nonumber\\
\left\langle \frac{\delta\mathbf{Y}\left(  \tau^{+}\right)  }{\delta V\left(
\tau\right)  }\right\rangle  &  =\mathcal{D}\left\langle \mathbf{Y}\left(
\tau\right)  \right\rangle , \label{Evolut for <dY/dh> WN}%
\end{align}
which coincides with Eq. (\ref{Evolut for <Y> in WN}), obtained for
$\left\langle \mathbf{Y}\left(  t\right)  \right\rangle $ in the white noise
model (Appendix \ref{Sect: GLE for WN}). The solution of Eq.
(\ref{Evolut for <dY/dh> WN}) is%
\begin{equation}
\left\langle \frac{\delta\mathbf{Y}\left(  t\right)  }{\delta V\left(
\tau\right)  }\right\rangle =\left\langle \mathcal{K}_{wn}\left(
t-\tau\right)  \right\rangle \mathcal{D}\left\langle \mathbf{Y}\left(
\tau\right)  \right\rangle , \label{K for WN}%
\end{equation}
where $\left\langle \mathcal{K}_{wn}\left(  t-\tau\right)  \right\rangle
\equiv e^{\mathcal{M}\left(  \epsilon,g\right)  \left(  t-\tau\right)  }$ and
matrix $\mathcal{M}\left(  \epsilon,g\right)  =\mathcal{C}+\frac{g}%
{2}\mathcal{D}^{2}$ is given in Eq. (\ref{M-matrix}). Thus, decoupling
(\ref{Decoupling}) is obtained automatically within the white noise
approximation, whereas the mean propagator in (\ref{Decoupling}) is
approximated by the exponent%
\begin{equation}
\left\langle \mathcal{K}\left(  t\right)  \right\rangle \approx\left\langle
\mathcal{K}_{wn}\left(  t\right)  \right\rangle \equiv e^{\mathcal{M}\left(
\epsilon,g\right)  t}. \label{K for WN gen}%
\end{equation}
For weak disorder,\ using approximation $\mathcal{M}\left(  \epsilon,g\right)
\approx\mathcal{C}$ in the above expression for $\left\langle \mathcal{K}%
_{wn}\left(  t\right)  \right\rangle $, and substituting it into Eq.
(\ref{EigVal eq for Approxs}), one recovers the Born approximation
(\ref{Born approx}), which is valid for an arbitrary $R_{c}$. It follows that
validity condition $kR_{c}\ll1$ is relaxed in the limit of weak disorder
(while the second one, $\Lambda R_{c}\ll1$, is usually fulfilled).

Substitution of Eq. (\ref{K for WN gen}) into the eigenvalue equation
(\ref{EigVal eq for Approxs}) yields an integral $\int_{0}^{\infty}%
C_{2}\left(  \tau\right)  e^{\mathcal{M}\left(  \epsilon,g\right)  \tau
}e^{-4\Lambda\tau}d\tau$, which is calculated by diagonalizing matrix
$\mathcal{M}\left(  \epsilon,g\right)  $. The latter is not a normal matrix
($\mathcal{MM}^{\dagger}\neq\mathcal{M}^{\dagger}\mathcal{M}$), and can be
written as \cite{Horn}%
\begin{equation}
\mathcal{Q}_{L}\mathcal{M}\left(  \epsilon,g\right)  \mathcal{Q}%
_{R}=\mathcal{\tilde{M}}\left(  \epsilon,g\right)  =4\left[
\begin{array}
[c]{ccc}%
\Lambda_{1} & 0 & 0\\
0 & \Lambda_{2} & 0\\
0 & 0 & \Lambda_{3}%
\end{array}
\right]  , \label{Diagonalization}%
\end{equation}
where eigenvalues $\Lambda_{i}=\Lambda_{i}\left(  \epsilon,g\right)  $ are
given in Eq. (\ref{Lambda_i}), while matrices $\mathcal{Q}_{L}=\left(
u_{1}^{L},u_{2}^{L},u_{3}^{L}\right)  ^{\dagger}$ and $\mathcal{Q}_{R}=\left(
u_{1}^{R},u_{2}^{R},u_{3}^{R}\right)  =\mathcal{Q}_{L}^{-1}$ are composed of
the left and the right eigenvectors of $\mathcal{M}\left(  \epsilon,g\right)
$, given in Eq.\ (\ref{Left+Right EigVecs WN}). Inserting the white noise
approximation (\ref{K for WN gen}) into Eq. (\ref{EigVal eq for Approxs}) and
using parametrization of the correlation function, Eqs.
(\ref{eq. DO properties}) and (\ref{eq. DO parameter}), one obtains%
\begin{equation}
\det\left[  \mathcal{C}+\frac{g}{2}\mathcal{DQ}_{R}\mathcal{HQ}_{L}%
\mathcal{D}-4\tilde{\Lambda}\right]  =0, \label{Det.1}%
\end{equation}
where the diagonal matrix%
\begin{equation}
\mathcal{H}=h\left(  R_{c}\left[  4\Lambda\mathbf{-}\mathcal{\tilde{M}}\left(
\epsilon,g\right)  \right]  \right)  ,
\end{equation}
is expressed in terms of the Laplace transform of the dimensionless
correlation function%
\begin{equation}
h\left(  z\right)  =\int_{0}^{\infty}\Gamma\left(  s\right)  e^{-zs}ds,\ z\in%
\mathbb{C}
. \label{h function}%
\end{equation}
Note that, if the dimensionless correlation function $\Gamma\left(  s\right)
$ decays slower than exponentially, then $h\left(  z\right)  $ is defined only
for $\operatorname{Re}z\geq0$. As a consequence, the small-$R_{c}$
approximation will turn out to be inapplicable in the case of slowly-decaying
(e.g. by power-law) correlation functions. This case should be treated with
some care and will be discussed later in this section.

From now on, we restrict ourselves to the case $\epsilon>-\frac{3}{4}g^{2/3}$.
In this regime, eigenvalues of $\mathcal{M}\left(  \epsilon,g\right)  $
satisfy $\Lambda_{3}=\Lambda_{2}^{\ast}$ (cf. (\ref{Lambda_i})), and matrix
$\mathcal{H}$ can be written as%
\begin{equation}
\mathcal{H}=%
\begin{bmatrix}
h_{1} & 0 & 0\\
0 & h_{r}+ih_{i} & 0\\
0 & 0 & h_{r}-ih_{i}%
\end{bmatrix}
, \label{H matrix (2)}%
\end{equation}
where%
\begin{gather}
h_{1}=h\left(  4R_{c}\left[  \tilde{\Lambda}\mathbf{-}\Lambda_{1}\right]
\right)  ,\ h_{r}=\operatorname{Re}h\left(  4R_{c}\left[  \tilde{\Lambda
}\mathbf{-}\Lambda_{2}\right]  \right)  ,\nonumber\\
h_{i}=\operatorname{Im}h\left(  4R_{c}\left[  \tilde{\Lambda}\mathbf{-}%
\Lambda_{2}\right]  \right)  , \label{h_funcs}%
\end{gather}
and $h\left(  z\right)  $ is defined in Eq. (\ref{h function}). Then, the
matrix product in Eq. (\ref{Det.1}) becomes%
\begin{equation}
\mathcal{DQ}_{R}\mathcal{HQ}_{L}\mathcal{D=}\frac{1}{12\Lambda_{1}%
^{2}+\epsilon}%
\begin{bmatrix}
0 & 0 & 0\\
f_{12} & f_{22} & 0\\
f_{13} & f_{12} & 0
\end{bmatrix}
, \label{M_prod_2.2}%
\end{equation}
where%
\begin{align}
f_{12}  &  =2^{3/2}\Lambda_{1}\left(  h_{1}-h_{r}\right)  +\sqrt{2}%
\frac{6\Lambda_{1}^{2}+\epsilon}{\sqrt{3\Lambda_{1}^{2}+\epsilon}}%
h_{i},\nonumber\\
f_{22}  &  =h_{1}-h_{r}-\frac{3\Lambda_{1}}{\sqrt{3\Lambda_{1}^{2}+\epsilon}%
}h_{i},\nonumber\\
f_{13}  &  =8\Lambda_{1}^{2}\left(  h_{1}+2h_{r}\right)  +2\epsilon
h_{r}+\frac{2\Lambda_{1}\epsilon h_{i}}{\sqrt{3\Lambda_{1}^{2}+\epsilon}}.
\label{M_prod_2.2_elems}%
\end{align}
Finally, inserting matrix (\ref{M_prod_2.2}) into the eigenvalue equation
(\ref{Det.1})%
\begin{equation}
\det\left[  \mathcal{C}+\frac{g}{2}\mathcal{DQ}_{R}\mathcal{HQ}_{L}%
\mathcal{D}-4\tilde{\Lambda}\right]  =0, \label{Det.2}%
\end{equation}
one calculates the generalized LE $\Lambda$ as the largest real-valued root of
the determinant.

The roots of the corresponding non-linear equation can only be found
numerically. Simple analytical expressions, however, can be obtained in some
limiting cases. In weak disorder, $\Lambda_{1}^{2}\ll\epsilon$, one recovers
the Born approximation (\ref{Born approx}), as already discussed after Eq.
(\ref{K for WN gen}). In sufficiently strong disorder, such that $\Lambda
_{1}^{2}\gg\epsilon$, one can discard $\epsilon$ in Eqs.
(\ref{M_prod_2.2_elems}), which simplifies the expressions. The resulting
eigenvalue equation can be solved analytically in the limits $\Lambda_{1}%
R_{c}\ll1$ and $\Lambda_{1}R_{c}\gg1$. The first condition, $\Lambda_{1}%
R_{c}\ll1$, corresponds to the white noise limit and yields $\Lambda
\equiv\Lambda_{1}$, as expected. In the second limit, $\Lambda_{1}R_{c}\gg1$,
which is beyond validity of the small-$R_{c}$ approximation and is considered
formally, one obtains $\Lambda\approx\Lambda_{1}$ (more specifically
$1<\Lambda/\Lambda_{1}<1.16$). Between these two limits of small and large
$\Lambda_{1}R_{c}$, one still expects $\Lambda\approx\Lambda_{1}$, which was
also found solving the eigenvalue equation numerically for the correlation
functions listed in Table \ref{Table}. Thus, in strong disorder, $\Lambda
_{1}^{2}\gg\epsilon$, we obtain $\Lambda\approx\Lambda_{1}$ irrespectively of
the form of the disorder correlation function. This result was also obtained
in semiclassical approximation and in numerical simulations discussed below.
According to Eqs. (\ref{GLE in WN implicite}) and (\ref{eq. DO parameter}),
the above strong disorder condition can be reexpressed as $\left\vert
\epsilon\right\vert ^{3/2}\ll V_{0}^{2}R_{c}$.%

\begin{table}[tbp] \centering
\begin{tabular}
[c]{|c|c|c|}\hline
Correlation & $\Gamma\left(  x\right)  $ & $h\left(  z\right)  $\\\hline
Exponential & $e^{-x}$ & $\left(  1+z\right)  ^{-1}$\\
Gaussian & $e^{-\pi x^{2}/4}$ & $e^{z^{2}/\pi}\operatorname{erfc}\frac
{z}{\sqrt{\pi}}$\\
"Speckle" & $\ \operatorname{sinc}^{2}\frac{x}{2}\ $ & $\ \frac{2}{\pi}%
\arctan\frac{\pi}{z}-\frac{z}{\pi^{2}}\ln\frac{z^{2}+\pi^{2}}{z^{2}%
},\ \operatorname{Re}z\geq0\ $\\\hline
\end{tabular}
\caption{Dimesionless correlation function $\Gamma\left(x\right)$ and its Laplace transform
$h\left(z\right)$, Eq. (\ref{h function}), for the models of  disorder implemented in simulations.}\label{Table}%
\end{table}%

\subsection{A regularized white noise approximation}

In what follows, we refer to the approximation method described in the
previous subsection, Eqs. (\ref{M_prod_2.2})-(\ref{Det.2}), as the standard
white noise approximation (SWNA). As already noted after Eq. (\ref{Det.1}),
this method is not applicable to slowly decaying correlations. We now discuss
this point in some more detail and propose a simple regularization to extend
the applicability of the approximation.

The white noise propagator $\left\langle \mathcal{K}_{wn}\left(  t\right)
\right\rangle $, involved in approximation (\ref{K for WN}) for $\left\langle
\frac{\delta\mathbf{Y}\left(  t\right)  }{\delta V\left(  \tau\right)
}\right\rangle $, has the largest eigenvalue $e^{4\Lambda_{1}\left(
\epsilon,g\right)  t}$, where $\Lambda_{1}\left(  \epsilon,g\right)  $ is the
generalized LE for the uncorrelated disorder (see Appendix
\ref{Sect: GLE for WN}). As follows from the Born approximation
(\ref{Born approx}), it is usually larger than the generalized LE $\Lambda$
for the correlated potential with the same intensity $g$, at least for
sufficiently weak disorder. Therefore, if correlation function $C_{2}\left(
\tau\right)  $ decays slower than $e^{-4\left(  \Lambda_{1}\left(
\epsilon,g\right)  -\Lambda\right)  \tau}$, then the integral in the
eigenvalue equation (\ref{EigVal eq for Approxs}), which can be estimated as%
\begin{equation}
\int\limits_{0}^{\infty}C_{2}\left(  \tau\right)  \left\langle \mathcal{K}%
_{wn}\left(  t\right)  \right\rangle e^{-4\tilde{\Lambda}\tau}d\tau\sim
\int\limits_{0}^{\infty}C_{2}\left(  \tau\right)  e^{4\left(  \Lambda
_{1}\left(  \epsilon,g\right)  -\tilde{\Lambda}\right)  \tau}d\tau,
\label{Diverg-Int}%
\end{equation}
diverges for $\tilde{\Lambda}=\Lambda$. In the final expressions, Eqs.
(\ref{Det.1})-(\ref{M_prod_2.2_elems}), this potential divergence resides in
function $h_{1}=\int_{0}^{\infty}\Gamma\left(  \tau\right)  e^{4\left(
\Lambda_{1}\left(  \epsilon,g\right)  -\tilde{\Lambda}\right)  \tau}d\tau$,
defined by Eqs. (\ref{h function}) and (\ref{h_funcs}). Thus, for slowly, e.g.
sub-exponentially, decaying correlations, $\Lambda$ can not be a solution of
the eigenvalue equation and the standard white noise approximation is a priori
inapplicable. Divergence of the integral in Eq. (\ref{Diverg-Int}) as an
artifact of the white noise approximation (\ref{K for WN}), which yields a
wrong large-time asymptotics $\left\langle \frac{\delta\mathbf{Y}\left(
t\right)  }{\delta V\left(  \tau\right)  }\right\rangle \sim e^{4\Lambda
_{1}\left(  \epsilon,g\right)  (t-\tau)}$. Indeed, as noted after Eq.
(\ref{Evolut for dY/dh}), the true asymptotics of the moment of the functional
derivative is $\left\langle \frac{\delta\mathbf{Y}\left(  t\right)  }{\delta
V\left(  \tau\right)  }\right\rangle \sim e^{4\Lambda(t-\tau)}$, i.e. the same
as for $\left\langle \mathbf{Y}\left(  t\right)  \right\rangle $. Therefore,
if $\left\langle \mathcal{K}_{wn}\left(  t\right)  \right\rangle $ in
(\ref{Diverg-Int}) had a "correct" asymptotics, then the integral would
converge at $\tilde{\Lambda}=\Lambda$ as long as $\int_{0}^{\infty}%
C_{2}\left(  \tau\right)  d\tau\equiv\frac{g}{2}<\infty$. Thus, some kind of a
regularization is required for $\tau\gg R_{c}$. One possibility is to
introduce an upper cutoff of the order of $R_{c}$ into the integral for
$h_{1}$. Such cutoff was effectively used in Ref. \cite{Iomin-09} for another
scheme of a closure of Eq. (\ref{Evolut for <Y> with N-F}). A more simple
regularization is just to set $h_{1}=\int_{0}^{\infty}\Gamma\left(
\tau\right)  d\tau=1$, which amounts to a self-consistent modification of the
white noise propagator $\left\langle \mathcal{K}_{wn}\left(  t\right)
\right\rangle $ in the approximation (\ref{K for WN gen}) by replacing its
largest eigenvalue $e^{4\Lambda_{1}\left(  \epsilon,g\right)  t}$ with
$e^{4\Lambda t}$. We use the second approach and refer to it as a "regularized
white noise approximation" (RWNA). Thus, the regularized white noise
approximation is given by Eqs. (\ref{M_prod_2.2})-(\ref{Det.2}), with $h_{1}$
replaced by unity. Note, that this method can be regarded as applying the
decoupling approximation (\ref{Decoupling}) with the corresponding effective
propagator, different from that in Eq. (\ref{K for WN gen}).

\subsection{Numerical test of small-$R_{c}$
approximation\label{Sect: num tests}}

The standard and the regularized white noise approximations are compared to
the direct numerical simulation for the three types of disorder correlations
given in Table \ref{Table}. Analytical values of $\Lambda$ are obtained by
finding numerically the largest real-valued root of the determinant
(\ref{Det.1}). The Monte-Carlo simulations were done for the tight-binding
model (\ref{Tight-binding Shrod. Eq.}) with energy $E=-2\cos k$ near the band
edge, which translates to the continuous model (\ref{eqn1}) with a positive
energy $\epsilon=k^{2}$. Further details of the numerical simulations are
given in Appendix \ref{Sect: Numrics}.%

\begin{figure}
[ptb]
\begin{center}
\includegraphics[
height=5.3375in,
width=2.8908in
]%
{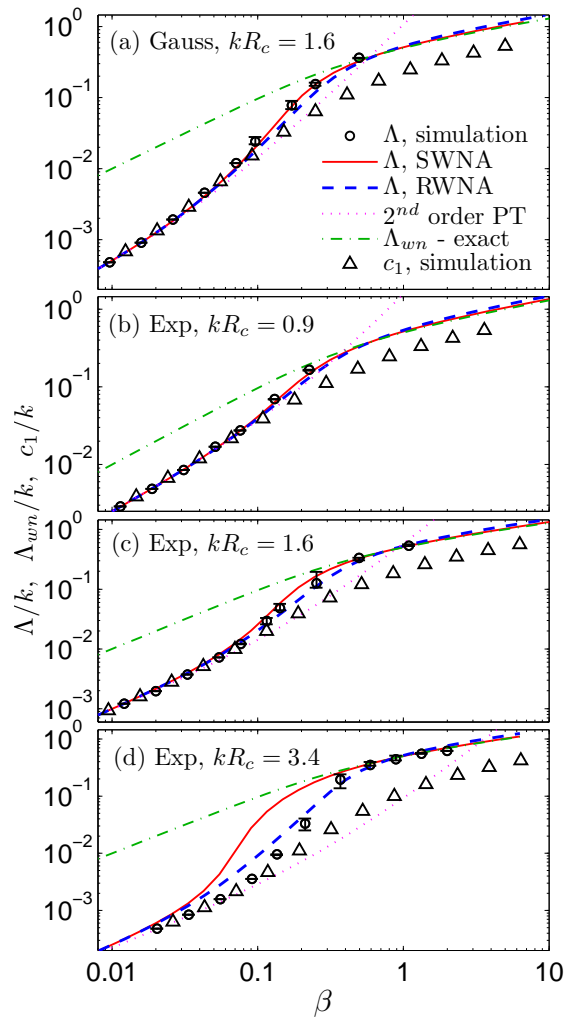}%
\caption{(color online) The standard (SWNA) and the regularized (RWNA) white
noise approximations for the generalized LE $\Lambda$ are compared to
numerical simulation for (a) Gaussian and (b-d) exponential correlation
functions (Table \ref{Table}). Legends in (a) apply to all panels. GLE,
normalized to the wavevector $k$, is plotted against the dimensionless
disorder intensity $\beta$, Eq. (\ref{beta-def}). Error bars indicate
uncertainty of the Monte-Carlo calculation of $\Lambda$. For comparison, we
also plot $2^{nd}$ order (i.e., one order beyond the Born approx.)
perturbative calculation of $\Lambda$ due to Ref. \cite{Tessieri-02}, and the
exact GLE $\Lambda_{wn}$ for $\delta$-correlated disorder given by
$\Lambda_{1}$ in Eq. (\ref{Lambda_i}). In addition, LE $c_{1}$, obtained by
numerical simulation, is plotted with triangles. As expected for Gaussian
disorder \cite{Gurevich}, $\Lambda$ and $c_{1}$ coincide in weak disorder as
long as $2^{nd}$ order perturbation theory is valid. In strong disorder,
$\beta\sim1$, GLE for correlated disorder coincides with $\Lambda_{wn}\left(
\beta\right)  $.}%
\label{Fig. 1}%
\end{center}
\end{figure}

First, we examine models of disorder with the quickly decaying exponential and
Gaussian correlation functions (Table \ref{Table}). The results are compared
to the Monte-Carlo simulation in Fig. \ref{Fig. 1}. The error bars indicate
uncertainty of the Monte-Carlo calculation of the generalized LE, which is
explained in Appendix \ref{Sect: Numrics}. the generalized LE $\Lambda$ is
plotted for different values of $kR_{c}$ as a function of the dimensionless
noise intensity%
\begin{equation}
\beta\equiv\frac{gk}{8\epsilon^{2}}=\frac{V_{0}^{2}kR_{c}}{4\epsilon^{2}},
\label{beta-def}%
\end{equation}
where $g$ is defined in Eq. (\ref{eq. DO parameter}). Note that $\beta
=\Lambda/k$ in the Born approximation (\ref{Born approx}) for the white noise disorder.

The standard and the regularized white noise approximations yield close
results for $kR_{c}\lesssim1$, while Difference between them increases with
$R_{c}$ and, e.g., for the exponential correlation with $kR_{c}=3.4$, it
becomes of the order of ten (Fig. \ref{Fig. 1}(d)). Comparison with the
simulation results shows that both approximations are good for $kR_{c}%
\lesssim1$ (Fig. \ref{Fig. 1}(a-c)), but deteriorate with further increase of
$kR_{c}$, as expected. For example, for the exponential correlation with
$kR_{c}=3.4$, the resul of the standard approximation and the simulated values
of $\Lambda$ differ already by a factor of ten at moderately strong disorder,
$\beta\sim0.1$ (Fig. \ref{Fig. 1}(d)). The regularized white noise
approximation seems to be somewhat better, though the improvement is inconclusive.

\begin{figure}
[ptb]
\begin{center}
\includegraphics[
height=3.3848in,
width=3.2171in
]%
{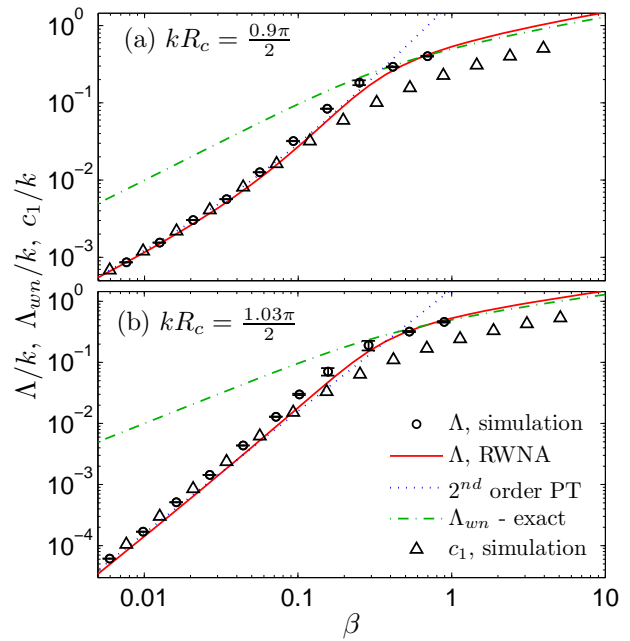}%
\caption{(color online) The regularized white noise approximation (RWNA) and
the numerical simulation of $\Lambda$ are compared for the power law "speckle"
correlation (See Table \ref{Table}) with (a) $kR_{c}=0.9\pi/2$ and (b)
$kR_{c}=1.03\pi/2$ (legends in (b) apply to both panels). Error bars indicate
uncertainty of the numerical calculation of $\Lambda$. Like in Fig.
\ref{Fig. 1}, $\Lambda_{wn}\left(  \beta\right)  $ and the $2^{nd}$ order
perturbative calculation of $\Lambda$ due to Ref. \cite{Tessieri-02} are given
for reference. Triangles represent numerical result for LE $c_{1}$. As in Fig.
\ref{Fig. 1}, $\Lambda$ coincides with $c_{1}$ in weak disorder, and with
$\Lambda_{wn}\left(  \beta\right)  $ in strong disorder.}%
\label{Fig. Speckle}%
\end{center}
\end{figure}

The main advantage of the regularized white noise approximation is its
applicability to the long-range correlations, as demonstrated in Fig.
\ref{Fig. Speckle}. As an example, we use "speckle" correlation function
$\Gamma\left(  x\right)  =\operatorname{sinc}^{2}\frac{x}{2}$ , where
$\operatorname{sinc}x\equiv\frac{\sin\pi x}{\pi x}$, which describes
correlation of the intensity in some quasi-one-dimensional laser speckle
patterns \cite{Goodman-84}. The latter are used to create disordered
potentials in experiments with cold atoms \cite{Aspect-Speckle-06}. Note,
however, that our Gaussian model does not describe a true speckle intensity,
which exhibits highly non-Gaussian fluctuations. This correlation function has
a remarkable property that the corresponding power spectrum, given by the
Fourier transform of the correlation function $2V_{0}^{2}\int_{0}^{\infty
}dx\Gamma\left(  x/R_{c}\right)  \cos qx$, is proportional to the "tent"
function $\left(  1-\left\vert qR_{c}/\pi\right\vert \right)  \Theta\left(
1-\left\vert qR_{c}/\pi\right\vert \right)  $ and vanishes for $\left\vert
qR_{c}\right\vert >\pi$. As a results, the Born approximation for both LE
$c_{1}$ and the generalized LE $\Lambda$, Eq. (\ref{Born approx}), vanishes
for $2kR_{c}>\pi$. In view of this fact, the regularized white noise
approximation is tested below and above this threshold for $2kR_{c}$ equal to
$0.9\pi$ and $1.03\pi$ respectively. In both cases we find agreement with the
simulation results within the factor of the order of unity (Fig.
\ref{Fig. Speckle}).%

Let us summarize the above comparison of the analytical and the numerical
results. The suggested simple regularization allows extension of the standard
white noise closure approximation to the cases of slowly decaying
correlations. In the regime of moderate disorder, $0.1\lesssim\beta\lesssim1$,
white noise approximation is applicable for $kR_{c}\lesssim1$ (we consider
$\epsilon=k^{2}>0$), in accord with the formal requirement $kR_{c}\ll1$ (cf.
Eq. (\ref{validity conds})). In the limit of weak disorder, $\beta\ll1$, our
approximation coincides with the perturbation theory (\ref{Born approx}), i.e.
becomes exact up to weak disorder corrections. Thus, for $\beta\ll1$, the
validity condition $kR_{c}\ll1$ is relaxed (cf. discussion after Eq.
(\ref{K for WN gen})). Finally, for strong disorder, $\beta>1$, numerical data
coincide with and, thus, confirm the analytical result $\Lambda\left(
\beta\right)  \approx\Lambda_{1}\left(  \beta\right)  \equiv\Lambda
_{wn}\left(  \beta\right)  $, obtained for $\left\vert \epsilon\right\vert
^{3/2}\ll V_{0}^{2}R_{c}$, i.e. $\beta\gg1$, at the end of Sec.
\ref{Sect: WN approx}. This is inspite of the fact that for $\beta>1$ in our
simulations we have $\Lambda R_{c}\gtrsim1$, which violates the white noise
approximation validity condition $\Lambda R_{c}\ll1$, Eq.
(\ref{validity conds}). In the next section, however, the same relation
$\Lambda\left(  \beta\right)  \approx\Lambda_{1}\left(  \beta\right)  $ is
obtained in the regime $\Lambda R_{c}\gg1$ \emph{specifically} for Gaussian disorder.

\section{Semiclassical regime of strong disorder\label{Sect: Semiclass}}

In this section we analytically calculate cumulant coefficients $c_{n}$,
defined by Eq. (\ref{Cn coefs-def}), in the regime when conditions
$\epsilon\,\,<V_{0}$ and $R_{c}\sqrt{V_{0}-\epsilon}\gg1$ are fulfilled, where
$V_{0}^{2}$ is the noise variance introduced in Eq. (\ref{eq. DO properties}).
An opposite regime, $R_{c}\sqrt{V_{0}-\epsilon}\ll1$, corresponds to the white
noise limit, for which the cumulants were calculated in Ref.
\cite{Schomerus-02}. We assume that $V_{0}$ and $R_{c}$ are typical height and
width of the random barriers, therefore probability of a tunneling through a
single typical barrier is exponentially small under the above conditions. In
this case, as argued in Ref. \cite{Deych-03}, transport is dominated by the
under-barrier tunneling and interference effects are negligible. Since $R_{c}$
is a spatial scale of the disorder $V\left(  x\right)  $, the second
condition, $R_{c}\sqrt{V_{0}-\epsilon}\gg1$, justifies the semiclassical
approximation for the under-barrier tunneling. Thus, an exponential
growth\ rate of the solution amplitude $A\left(  x\right)  =\sqrt{\psi
^{2}+\psi^{\prime2}}$ can be estimated as a magnitude of the imaginary part of
the semiclassical action:%
\begin{equation}
\ln\left\vert A\left(  L\right)  \right\vert =\int_{0}^{L}\kappa\left(
x\right)  dx. \label{QuasiCl approx}%
\end{equation}
Here $\kappa\left(  x\right)  =\Theta_{\epsilon}\left(  V\right)
\sqrt{V\left(  x\right)  -\epsilon}$ is an imaginary part of the complex
momentum, where $\Theta_{\epsilon}\left(  V\right)  \equiv\Theta\left(
V-\epsilon\right)  $ is a short notation for the unit step function. Function
$\Theta_{\epsilon}\left(  V\right)  $ selects regions where the energy is
below the disorder barriers and solution $\psi\left(  x\right)  $ grows
exponentially, while the remaining regions, where $\psi\left(  x\right)  $ has
an oscillatory behavior, are discarded.

Semiclassical approximation (\ref{QuasiCl approx}) directly relates
distribution of $\ln\left\vert A\left(  L\right)  \right\vert $ to the
statistical properties of the above-threshold excursions of the random process
$V\left(  x\right)  $. Disorder average of Eq. (\ref{QuasiCl approx}) yields
LE or the first cumulant coefficient%
\begin{equation}
c_{1}=\left\langle \kappa\right\rangle =\left\langle \kappa\right\rangle
_{\epsilon}\left\langle \Theta_{\epsilon}\left(  V\right)  \right\rangle ,
\label{Q-Cl_c1}%
\end{equation}
where $\left\langle \kappa\right\rangle _{\epsilon}\equiv\int_{\epsilon
}^{\infty}dV\sqrt{V-\epsilon}P_{\epsilon}\left(  V\right)  $ is a conditional
average for $V>\epsilon$ over the distribution $P_{\epsilon}\left(  V\right)
\equiv\Theta_{\epsilon}\left(  V\right)  P\left(  V\right)  /\int_{\epsilon
}^{\infty}dVP\left(  V\right)  $. Here $P\left(  V\right)  $ is distribution
of the disorder values at a single position.

Using semiclassical approximation (\ref{QuasiCl approx}), higher cumulant
coefficients $c_{n}$ can be expressed in terms of the corresponding joint
cumulants of $\kappa\left(  x\right)  $:%
\begin{equation}
c_{n}=\lim_{L\rightarrow\infty}\frac{1}{L}\int\limits_{0}^{L}\cdots
\int\limits_{0}^{L}%
{\displaystyle\prod_{i=1}^{n}}
dx_{i}\left\langle \left\langle
{\displaystyle\prod_{i=1}^{n}}
\kappa\left(  x_{i}\right)  \right\rangle \right\rangle . \label{Q-Cl_cum_1}%
\end{equation}
Since $V\left(  x\right)  $ is a stationary process, the joint cumulants
$\left\langle \left\langle \cdots\right\rangle \right\rangle $ in Eq.
(\ref{Q-Cl_cum_1}) depend only on the coordinate difference. The value of the
cumulants is maximal when all the coordinates coincide, and decays on the
scale $R_{c}$ with a distance between the points. This is because $R_{c}$ is
the only spatial scale in our model (\ref{eq. DO properties}). Thus, shifting
all the coordinates in the cumulant by $x_{n}$ and assuming $L\gg R_{c}$, the
integral in (\ref{Q-Cl_c1}) can be approximated as%
\begin{equation}
c_{n}\approx\int\limits_{-\infty}^{\infty}\cdots\int\limits_{-\infty}^{\infty}%
{\displaystyle\prod_{i=1}^{n-1}}
dx_{i}\left\langle \left\langle \kappa\left(  0\right)
{\displaystyle\prod_{i=1}^{n-1}}
\kappa\left(  x_{i}\right)  \right\rangle \right\rangle , \label{Q-Cl_cum_2}%
\end{equation}
where we have neglected the boundary effects at the corners $x_{1}%
=x_{2}=...=x_{n}=0,L$ and extended the integration to $\pm\infty$. For the
Gaussian process $V\left(  x\right)  $, an analytical calculation of the joint
cumulants of $\kappa\left(  x\right)  $ is quite involved. It is, however,
possible to obtain a simple estimate of the multiple integral in
(\ref{Q-Cl_cum_2}). As an example, consider the two-point correlator of the
regular step function $\Theta\left(  V\left(  x\right)  \right)  $, which can
be calculated analytically \cite{Blacknell-01}, and is given by%
\begin{equation}
\frac{\left\langle \left\langle \Theta\left(  V\left(  x\right)  \right)
\Theta\left(  V\left(  x^{\prime}\right)  \right)  \right\rangle \right\rangle
}{\left\langle \left\langle \Theta^{2}\left(  V\right)  \right\rangle
\right\rangle }=\frac{2}{\pi}\arcsin\left[  \Gamma\left(  \frac{x-x^{\prime}%
}{R_{c}}\right)  \right]  , \label{Theta_correl}%
\end{equation}
where the dimensionless correlation function $\Gamma\left(  x\right)  $ was
defined in Eq. (\ref{eq. DO properties}). As expected, the two-point
correlator of $\Theta\left(  V\left(  x\right)  \right)  $ decays in the same
manner as $\Gamma\left(  x\right)  $. Due to the normalization conditions
(\ref{normalization of Gamma2}), the integral of the expression in Eq.
(\ref{Theta_correl}) is of the order of $R_{c}$. Similarly, any $n$-order
cumulant in (\ref{Q-Cl_cum_2}) has maximum value $\left\langle \left\langle
\kappa^{n}\left(  0\right)  \right\rangle \right\rangle $ and decays with a
distance from the origin on the scale $R_{c}$. Therefore, rescaling the
integration variables in (\ref{Q-Cl_cum_2}) by $R_{c}$, the remaining
dimensionless integral can be grossly estimated as a volume of the $\left(
n-1\right)  $-dimensional unit sphere $S_{n-1}$. This gives the following
estimate%
\begin{equation}
c_{n}=B_{n}\left(  \frac{\epsilon}{V_{0}}\right)  S_{n-1}R_{c}^{n-1}%
\left\langle \left\langle \kappa^{n}\right\rangle \right\rangle ,
\label{Q-Cl_c_n}%
\end{equation}
where functions $B_{n}\left(  \epsilon/V_{0}\right)  $ are of the order of
unity (according to Eq. (\ref{Q-Cl_c1}), $B_{1}\equiv1$), and $\left\langle
\left\langle \kappa^{n}\right\rangle \right\rangle \equiv\left\langle
\left\langle \kappa^{n}\left(  x\right)  \right\rangle \right\rangle $, which
is independent of the position $x$. Coefficients $B_{n>1}\left(
\epsilon/V_{0}\right)  $ compensate for the approximation of the $\left(
n-1\right)  $-dimensional integral by the volume of sphere and depend weakly
on the specific form of the correlation function of $V\left(  x\right)  $ and
on the ratio $\epsilon/V_{0}$. Numerical values of $B_{n}\left(
\epsilon/V_{0}\right)  $ can be found in computer simulation by calculating
statistics of the quantity on the right hand side of Eq. (\ref{QuasiCl approx}%
), and fitting it to Eq. (\ref{Q-Cl_c_n}). As an example, we obtain
$B_{2}\left(  0\right)  \approx0.9$, $B_{2}\left(  \frac{1}{3}\right)
\approx0.8$ and $B_{3}\left(  0\right)  \approx1.6$, $B_{3}\left(  \frac{1}%
{3}\right)  \approx1.2$ for all three types of correlations given in Table
\ref{Table}.

Combining Eqs. (\ref{Q-Cl_c1}) and (\ref{Q-Cl_c_n}), one obtains the following
relations%
\begin{equation}
\frac{c_{n}}{c_{1}}=S_{n-1}B_{n}\left(  \frac{\epsilon}{V_{0}}\right)
f_{n}\left(  \frac{\epsilon}{V_{0}}\right)  \left(  c_{1}R_{c}\right)  ^{n-1},
\label{Q-Cl_scaling_gen}%
\end{equation}
where $f_{n}\left(  \frac{\epsilon}{V_{0}}\right)  \equiv\frac{\left\langle
\left\langle \kappa^{n}\right\rangle \right\rangle }{\left\langle
\kappa\right\rangle ^{n}}$ are dimensionless functions, whose specific form
depends only on the disorder distribution $P\left(  V\right)  $. This result
should be contrasted with the weak disorder relations (\ref{SPS rel}).

Using the fact that $\Theta_{\epsilon}\left(  V\right)  $ is either $0$ or
$1$, it is convenient to express cumulants of $\kappa\left(  x\right)  $ in
terms of $\left\langle \left\langle \Theta_{\epsilon}^{n}\left(  V\right)
\right\rangle \right\rangle $ and $\left\langle \left\langle \kappa
^{n}\right\rangle \right\rangle _{\epsilon}$, where $\left\langle \left\langle
\cdots\right\rangle \right\rangle _{\epsilon}$ denotes a "conditional
cumulant", calculated with distribution $P_{\epsilon}\left(  V\right)  $ [see
Appendix \ref{Sect: Auxiliary}]. For example, for $n=2,3$ one obtains
\begin{align}
f_{2}\left(  \frac{\epsilon}{V_{0}}\right)   &  =\frac{\left\langle
\left\langle \Theta_{\epsilon}^{2}\left(  V\right)  \right\rangle
\right\rangle }{\left\langle \Theta_{\epsilon}\left(  V\right)  \right\rangle
^{2}}+\frac{1}{\left\langle \Theta_{\epsilon}\left(  V\right)  \right\rangle
}\frac{\left\langle \left\langle \kappa^{2}\right\rangle \right\rangle
_{\epsilon}}{\left\langle \kappa\right\rangle _{\epsilon}^{2}},\nonumber\\
f_{3}\left(  \frac{\epsilon}{V_{0}}\right)   &  =\frac{1}{\left\langle
\Theta_{\epsilon}\left(  V\right)  \right\rangle ^{2}}\frac{\left\langle
\left\langle \kappa^{3}\right\rangle \right\rangle _{\epsilon}}{\left\langle
\kappa\right\rangle _{\epsilon}^{3}}+\nonumber\\
&  +3\frac{\left\langle \left\langle \Theta_{\epsilon}^{2}\left(  V\right)
\right\rangle \right\rangle }{\left\langle \Theta_{\epsilon}\left(  V\right)
\right\rangle ^{3}}\frac{\left\langle \left\langle \kappa^{2}\right\rangle
\right\rangle _{\epsilon}}{\left\langle \kappa\right\rangle _{\epsilon}^{2}%
}+\frac{\left\langle \left\langle \Theta_{\epsilon}^{3}\left(  V\right)
\right\rangle \right\rangle }{\left\langle \Theta_{\epsilon}\left(  V\right)
\right\rangle ^{3}} \label{Q-Cl_c_2&3}%
\end{align}
This expansion is helpful, since it shows separate contributions of the
fluctuation of the indicator function $\Theta_{\epsilon}\left(  V\right)  $
and of the barrier height fluctuation above the level $\epsilon$.

For the Gaussian distribution, $P\left(  V\right)  =\left(  2\pi V_{0}%
^{2}\right)  ^{-1/2}e^{-V^{2}/2V_{0}^{2}}$, explicit expression for $c_{1}$,
$f_{2}\left(  \frac{\epsilon}{V_{0}}\right)  $ and $f_{3}\left(
\frac{\epsilon}{V_{0}}\right)  $ are given in Eqs. (\ref{Q-Cl_c1(App)}) and
(\ref{Q-Cl_f_n(App)}).

Finally, the semiclassical approximation (\ref{QuasiCl approx}) can be used to
calculate the generalized LE $\Lambda$ in strong disorder limit%
\begin{equation}
\Lambda_{sc}=\lim_{L\rightarrow\infty}\frac{1}{4L}\ln\left\langle \exp\left[
2\int_{0}^{L}\kappa\left(  x\right)  dx\right]  \right\rangle ,
\label{QC_GLE_def}%
\end{equation}
where the disorder average of the exponential involves functional integration
over $V\left(  x\right)  $. As an example, we calculate $\Lambda_{sc}$ for
Gaussian disorder in the limit $\left\vert \epsilon\right\vert \ll V_{0}$ and
$\sqrt{V_{0}}R_{c}\gg1$, which justifies the stationary point approximation.
The latter yields%
\begin{equation}
\Lambda_{sc}=\frac{3}{8}\left(  2R_{c}V_{0}^{2}\right)  ^{1/3}.
\label{QC_GLE_res}%
\end{equation}
According to Eqs. (\ref{QC_GLE_res}) and (\ref{Q-Cl_c1(App)}), under the
conditions $\left\vert \epsilon\right\vert \ll V_{0}$ and $\sqrt{V_{0}}%
R_{c}\gg1$, LE ratio $\rho=\Lambda/c_{1}$ scales like $\left(  R_{c}V_{0}%
^{2}\right)  ^{1/3}/V_{0}^{1/2}\sim\left(  c_{1}R_{c}\right)  ^{1/3}$. For
comparison, the Gaussian part of $\rho$, i.e. $\rho_{G}=\frac{1}{2}\left(
1+\frac{c_{2}}{c_{1}}\right)  $ [cf. Eq.(\ref{Ro_def})], scales in this regime
as $c_{2}/c_{1}\sim c_{1}R_{c}$. The obtained scaling is closely related to
the statistics of the disorder fluctuations and, thus, is specific to the
considered Gaussian models (e.g., for strong binary disorder one expects
$\rho\sim const$).

Let us note that $\Lambda_{sc}$ in Eq. (\ref{QC_GLE_res}) matches very closely
the strong disorder limit of the small-$R_{c}$ approximation for $\Lambda$
(Sec. \ref{Sect: WN approx}, last paragraph). The latter reads $1<\Lambda
/\Lambda_{1}\left(  g\right)  <1.16$, where $g=2V_{0}^{2}R_{c}$ and
$\Lambda_{1}\left(  g\right)  =\frac{1}{2}\left(  \frac{g}{4}\right)  ^{1/3}$
is the generalized LE for $\delta$-correlated disorder in strong disorder
limit (cf. Eq.(\ref{GLE in WN implicite})). The semiclassical result
(\ref{QC_GLE_res}) gives $\Lambda_{sc}/\Lambda_{1}\left(  g\right)  =1.19$,
which is remarkably close to the small-$R_{c}$ approximation. Note that the
considered limit, $\sqrt{V_{0}}R_{c}\gg1$, implies that $\Lambda R_{c}\gg1$,
which is formally beyond validity of the small-$R_{c}$ approximation.%

\begin{figure}[ptb]\centering
\begin{tabular}
[c]{l}%
{\includegraphics[
height=2.3097in,
width=3.2146in
]%
{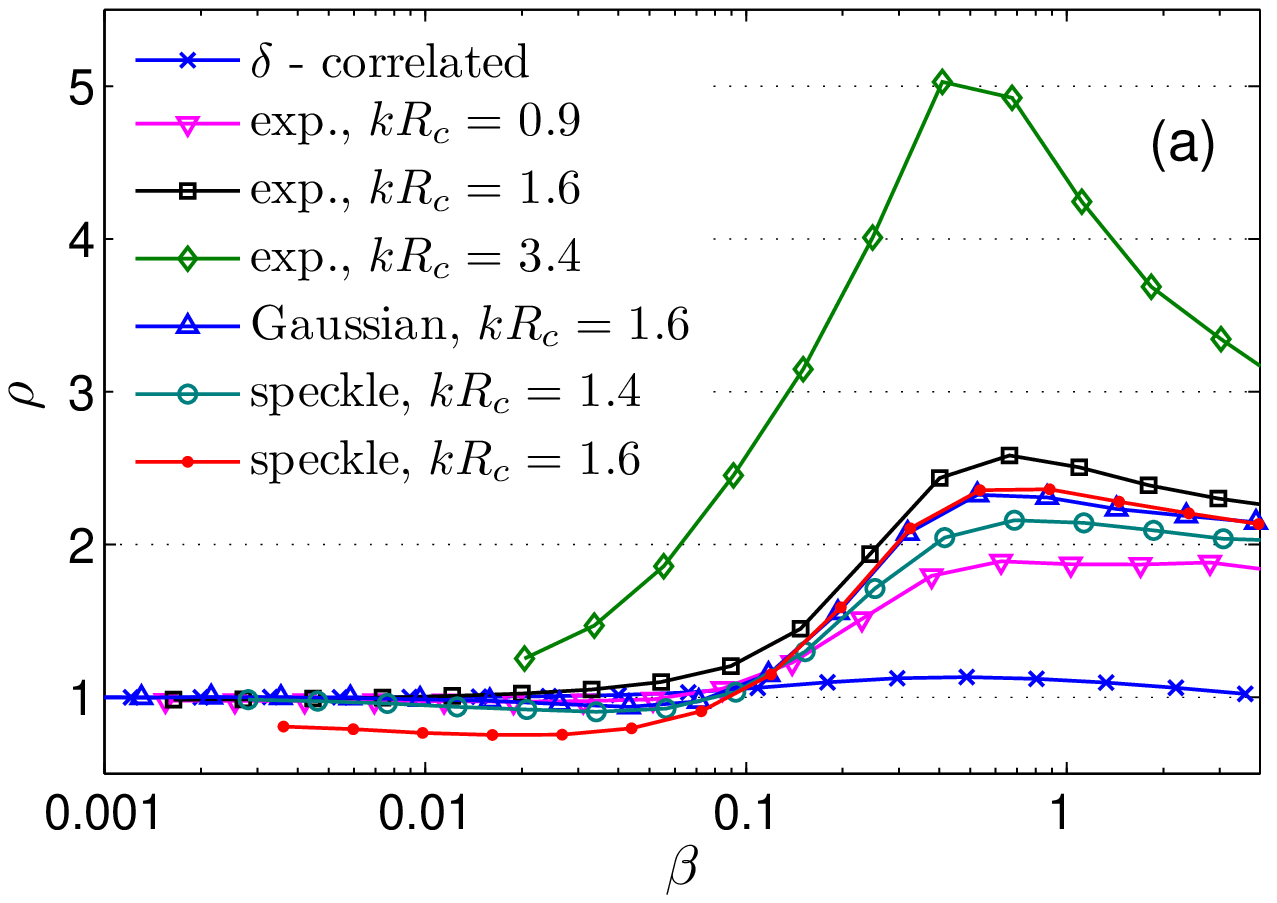}%
}%
\\%
\raisebox{-0.2167in}{\includegraphics[
height=2.2549in,
width=3.2154in
]%
{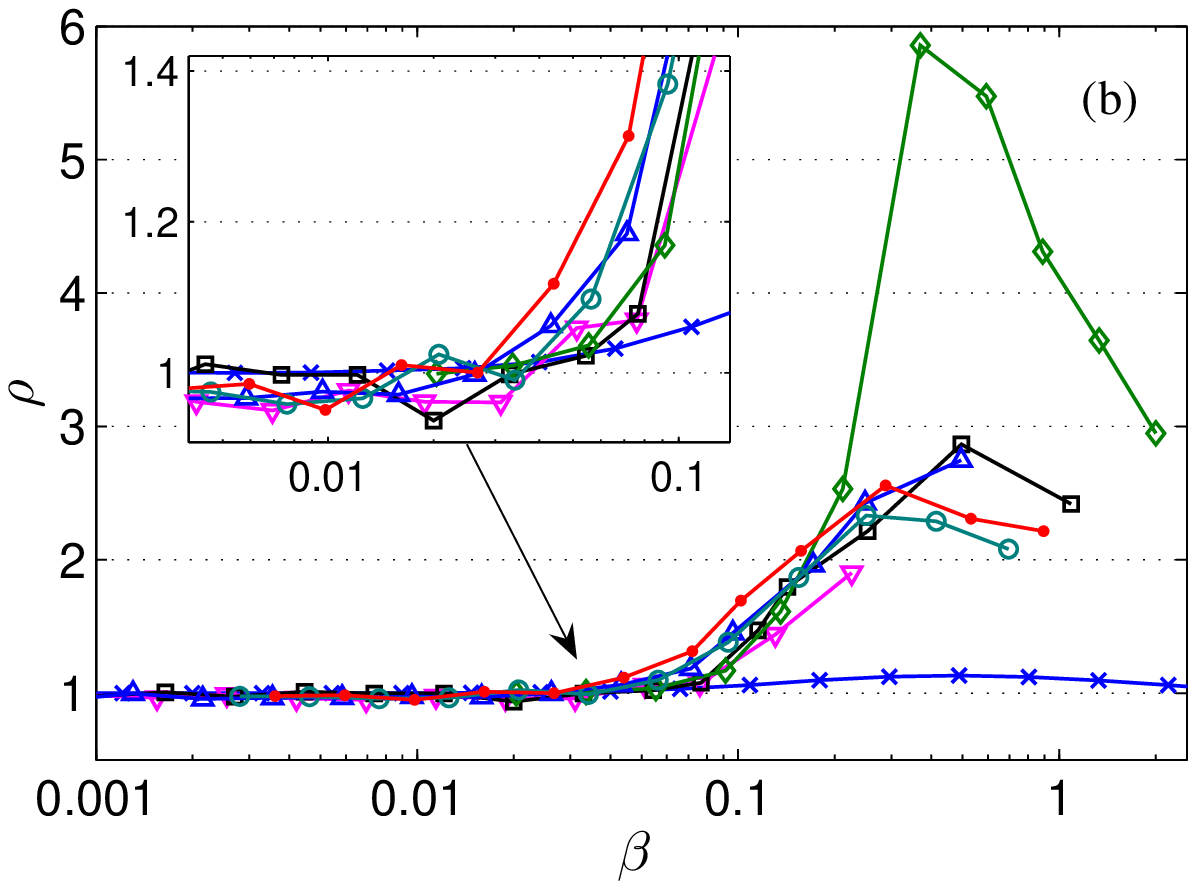}%
}%
\end{tabular}%
\caption{(color online) LE ratio $\rho=\Lambda
/c_1$ as a function of the dimensionless disorder intensity $\beta
$ [Eq. (\ref{beta-def}%
)]. LE $c_1$ is obtained by numerical simulations. GLE $\Lambda$ is
calculated using both the analytical  RWNA method, panel (a), as well as numerical simulations, panel (b).
Legends in (a) apply to both panels and specify type of correlation and value of $k R_c$.
The analytical and the numerical results complement each other in different regimes of disorder.
The numerics is noisy or unavailable at sufficiently large disorder intensities, $\beta
> 0.1$, where
analytical results demonstrate how $\rho$ increases with $k R_c$.
On the other hand, numerical results are more accurate for smaller disorder intensities, $\beta
\lesssim0.1$, where deviation from the limiting weak disorder value
$\rho= 1$ begins to develop, as shown in the inset of panel (b).
}%
\label{Fig. Res_gle}%
\end{figure}%
%

\begin{figure}[ptb]\centering
\begin{tabular}
[c]{l}%
{\includegraphics[
height=2.0789in,
width=3.2395in
]%
{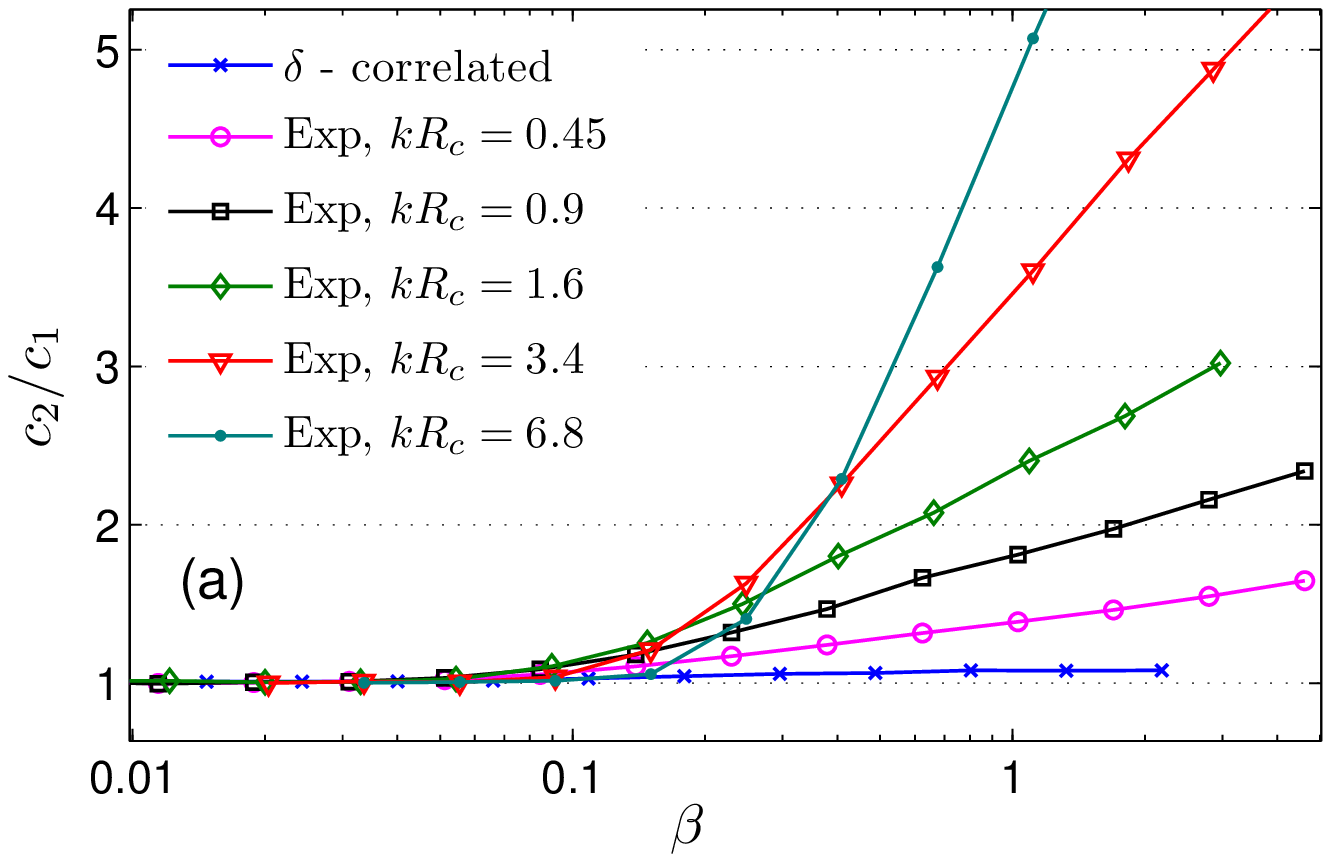}%
}%
\\%
{\includegraphics[
height=2.0714in,
width=3.2378in
]%
{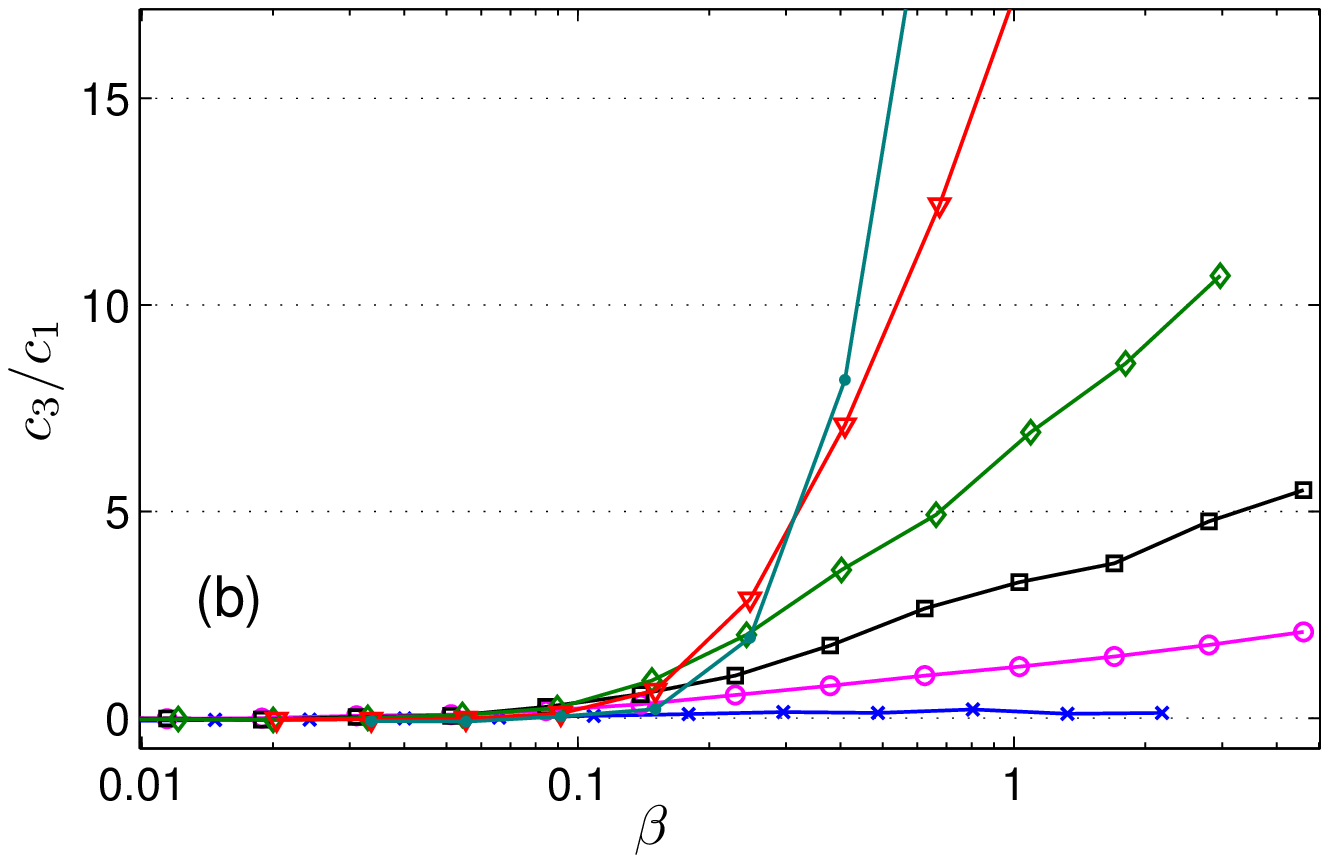}%
}%
\end{tabular}%
\caption
{(color online) Cumulant ratios $c_2/c_1$ (upper panel) and $c_3/c_1$ (lower panel) for the $\delta
$-correlated
and the exponentially corelated disorder plotted as a function of $\beta
$ (other models of correlation given in Table \ref{Table}
exhibit similar behavior, cf. Fig. \ref{Fig. Scaling_1}%
). Legends in (a)  apply to both panels
and specify the types of correlation and the values of $k R_c$.
}%
\label{Fig. Res_cums}%
\end{figure}%

\section{Transmission Statistics\label{Sect: Numerical results}}

In this section we discuss properties of the transmission coefficient
distribution in terms of the dimensionless ratios $c_{2}/c_{1}$, $c_{3}/c_{1}$
and $\rho=\Lambda/c_{1}$, Eqs. (\ref{T_cums_dimless}) and (\ref{Ro_def}). The
cumulant coefficients are simulated numerically for three types of
correlations listed in Table \ref{Table} and for the white noise model as
well. For the generalized LE $\Lambda$ we use both analytical and numerical
results. Details of the numerical simulations are given in Appendix
\ref{Sect: Numrics}. All calculations are performed at the same fixed value of
energy, $\epsilon=k^{2}>0$, and for different values of the correlation radius
$R_{c}$ and the disorder variance $V_{0}^{2}$. Disorder strength is
conveniently characterized by the dimensionless disorder intensity
$\beta=V_{0}^{2}kR_{c}/4\epsilon^{2}$, introduced in Eq. (\ref{beta-def}). For
each considered value of $R_{c}$, quantities $\Lambda$ and $c_{n}$ are studied
as a function of $V_{0}$ in the range from weak ($\beta\ll1$) to strong
($\beta\sim1$) disorder.

In Figs. \ref{Fig. Res_gle} and \ref{Fig. Res_cums} we plot $\rho$ and
cumulant ratios $c_{2}/c_{1}$ and $c_{3}/c_{1}$ respectively as a function of
$\beta$. Different data series correspond to different models of correlation
and different fixed values of $kR_{c}$, as indicated in the legends. All three
quantities, $\rho$, $c_{2}/c_{1}$ and $c_{3}/c_{1}$, exhibit qualitatively
similar behavior in weak and moderate disorder ($\beta<1$). As expected from
the results on the $\delta$-correlated disorder \cite{Pikovsky-03}%
,\cite{Schomerus-02}, universal relations $\rho=1$, $c_{2}/c_{1}=1$ and
$c_{3}/c_{1}=0$ (cf. Eqs.(\ref{SPS rel}) and (\ref{Ro_def})) are violated
beyond weak disorder. Our calculations show that deviation from these weak
disorder values is strongly enhanced in the presence of correlations and
increases with the correlation radius $R_{c}$. Namely, while this deviation is
small compared to unity for $\delta$-correlated disorder, it becomes of the
order of unity, or even larger, in the presence of correlations. Below we
discuss the obtained results and the corresponding parametric dependences.%

\begin{figure}[ptb]\centering
\begin{tabular}
[c]{l}%
{\includegraphics[
height=2.3113in,
width=3.2179in
]%
{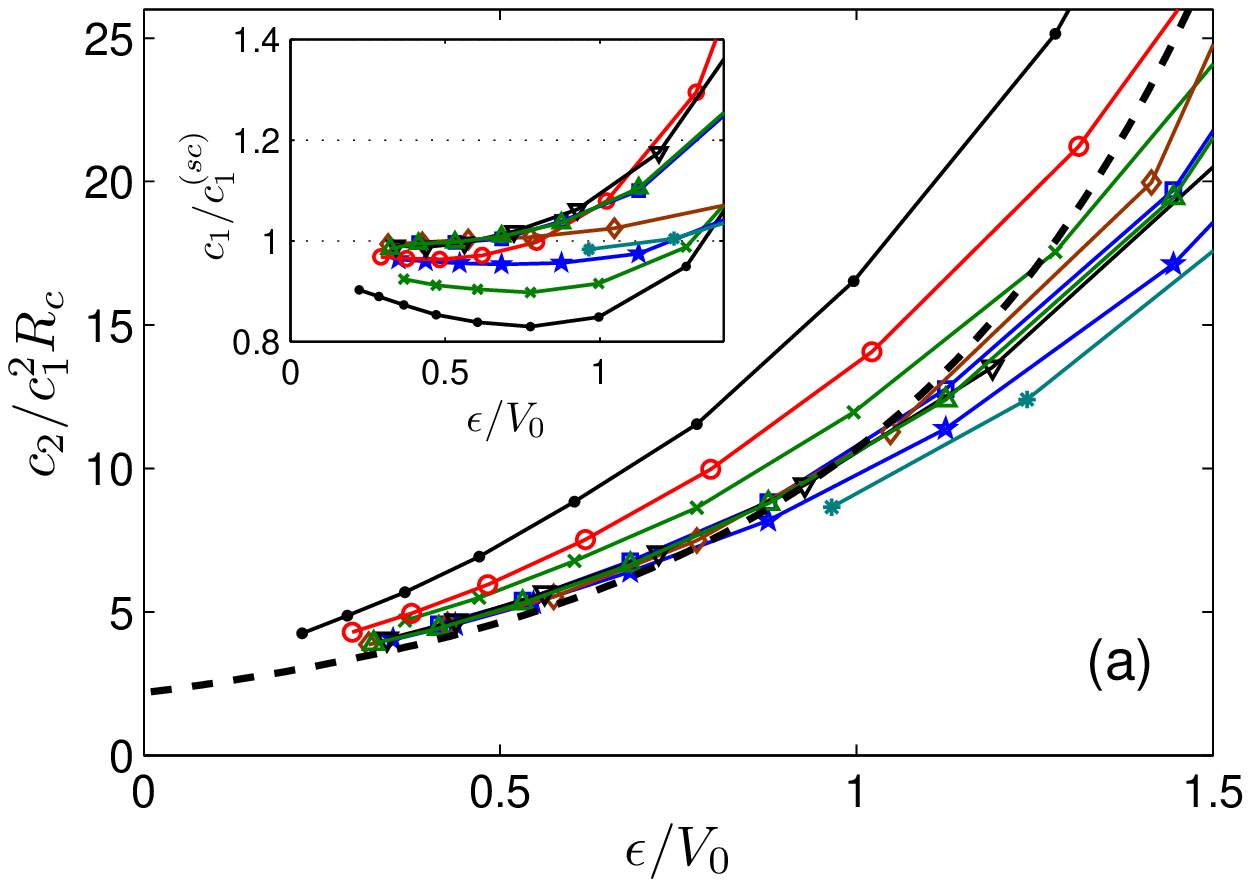}%
}%
\\%
{\includegraphics[
height=2.2482in,
width=3.2038in
]%
{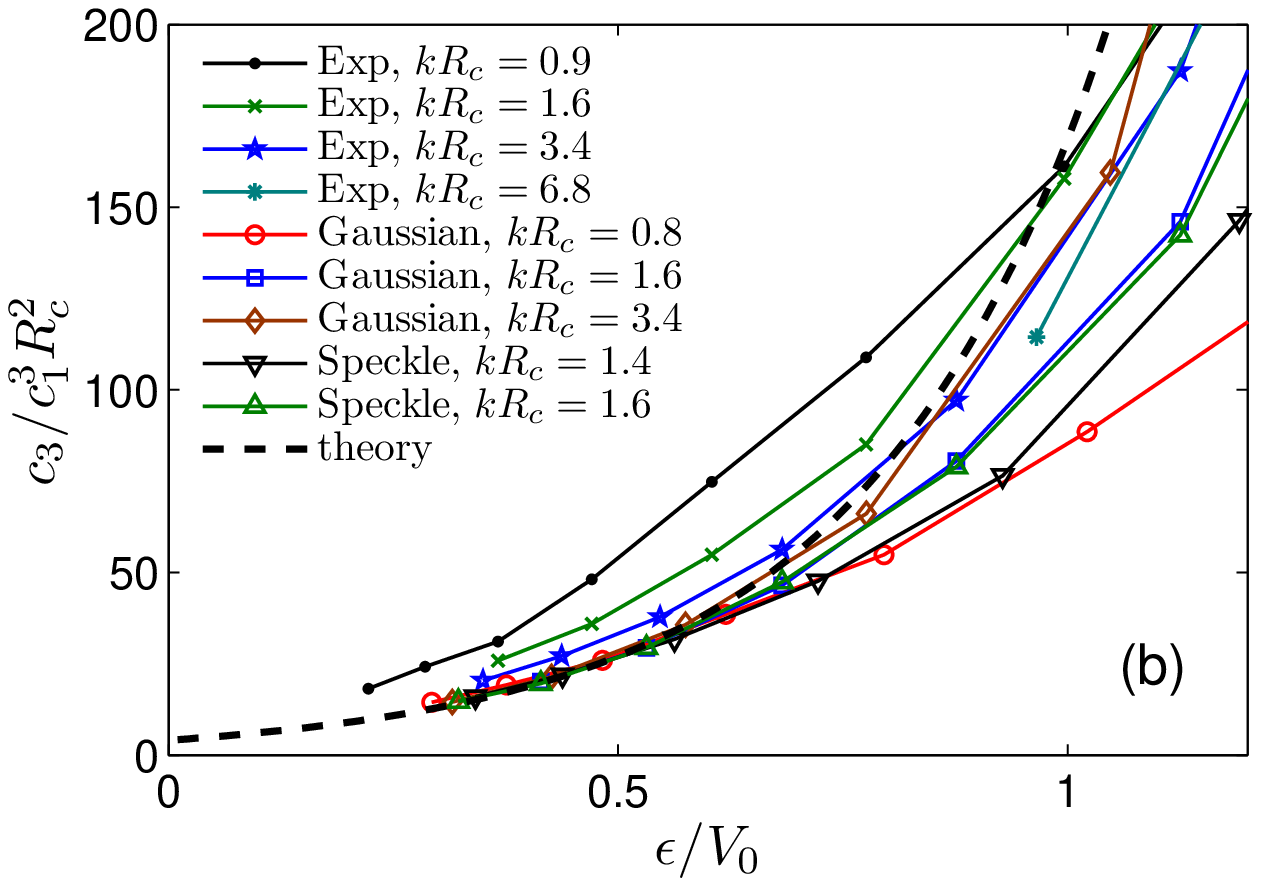}%
}%
\end{tabular}%
\caption{(color online) Semiclassical scaling relations (\ref{Q-Cl_c_2&3}%
) are verified
for the cumulant coeffcients $c_2$ (upper panel) and $c_3$ (lower panel) by plotting simulated
values of $c_n/c_1^n R_c^(n-1)$
as a function of the parameter $\epsilon
/V_0$. Legends in (b) apply to both panels.
The dashed lines represent analytical result due to Eq. (\ref{Q-Cl_c_2&3}%
) with
$B_2=0.8$ and $B_3=1.2$ (see text after Eq. (\ref{Q-Cl_c_2&3}%
)). To check the semiclassical approximation
for LE $c_1$, Eq. (\ref{Q-Cl_c1}%
), inset in panel (a) shows that ratio of the simulated value
of $c_1$ to its semiclassical approximation, denoted as $c_1^{sc}%
$, approaches unity for $\epsilon/V_0<1$. As expected,
semiclassical approximation improves for larger values of $R_c$ and smaller values of $\epsilon
/V_0$.
}%
\label{Fig. Scaling_2}%
\end{figure}%
%

\begin{figure}[ptb]\centering
\begin{tabular}
[c]{l}%
{\includegraphics[
height=2.4325in,
width=3.2179in
]%
{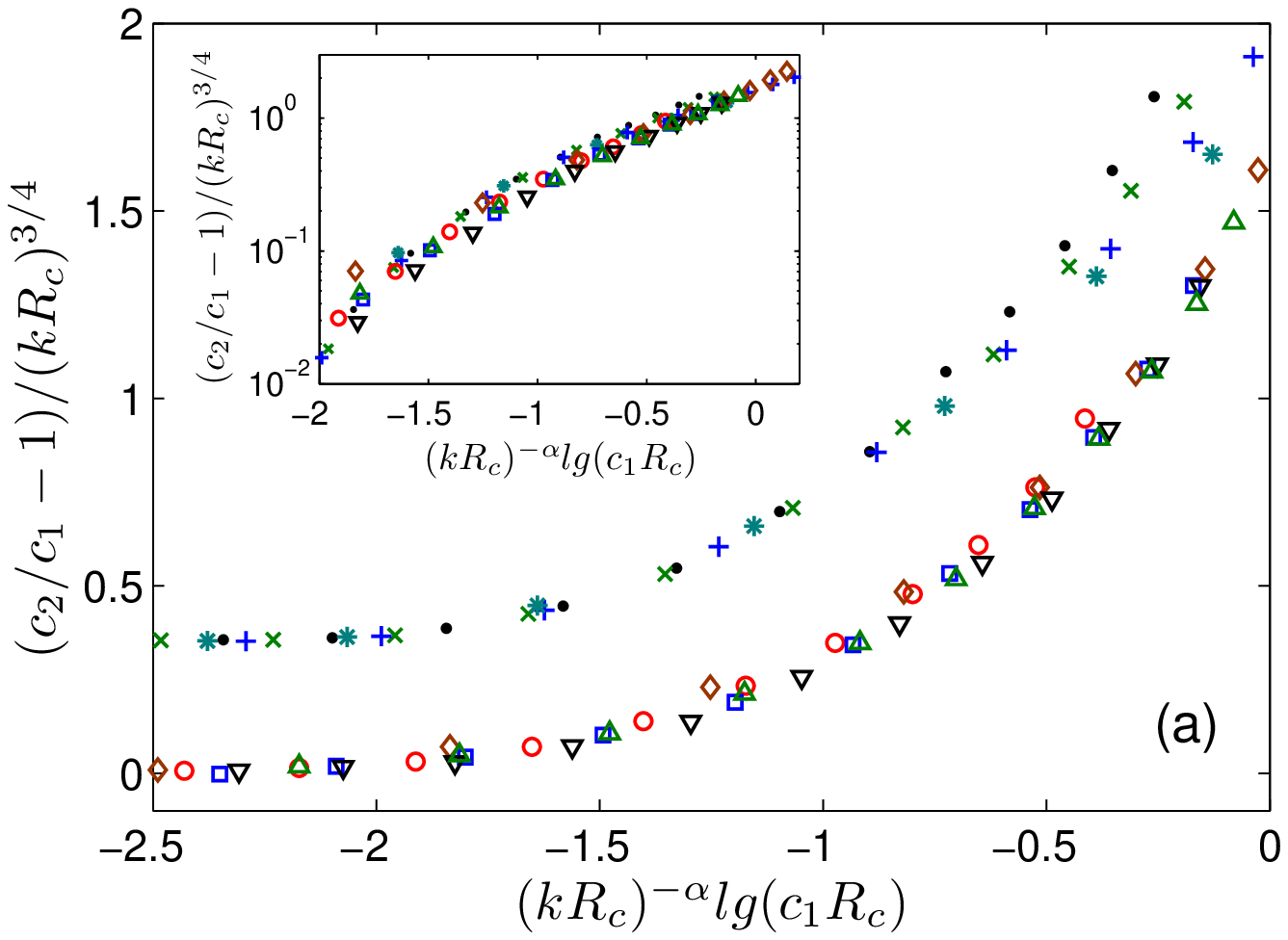}%
}%
\\%
{\includegraphics[
height=2.4583in,
width=3.2179in
]%
{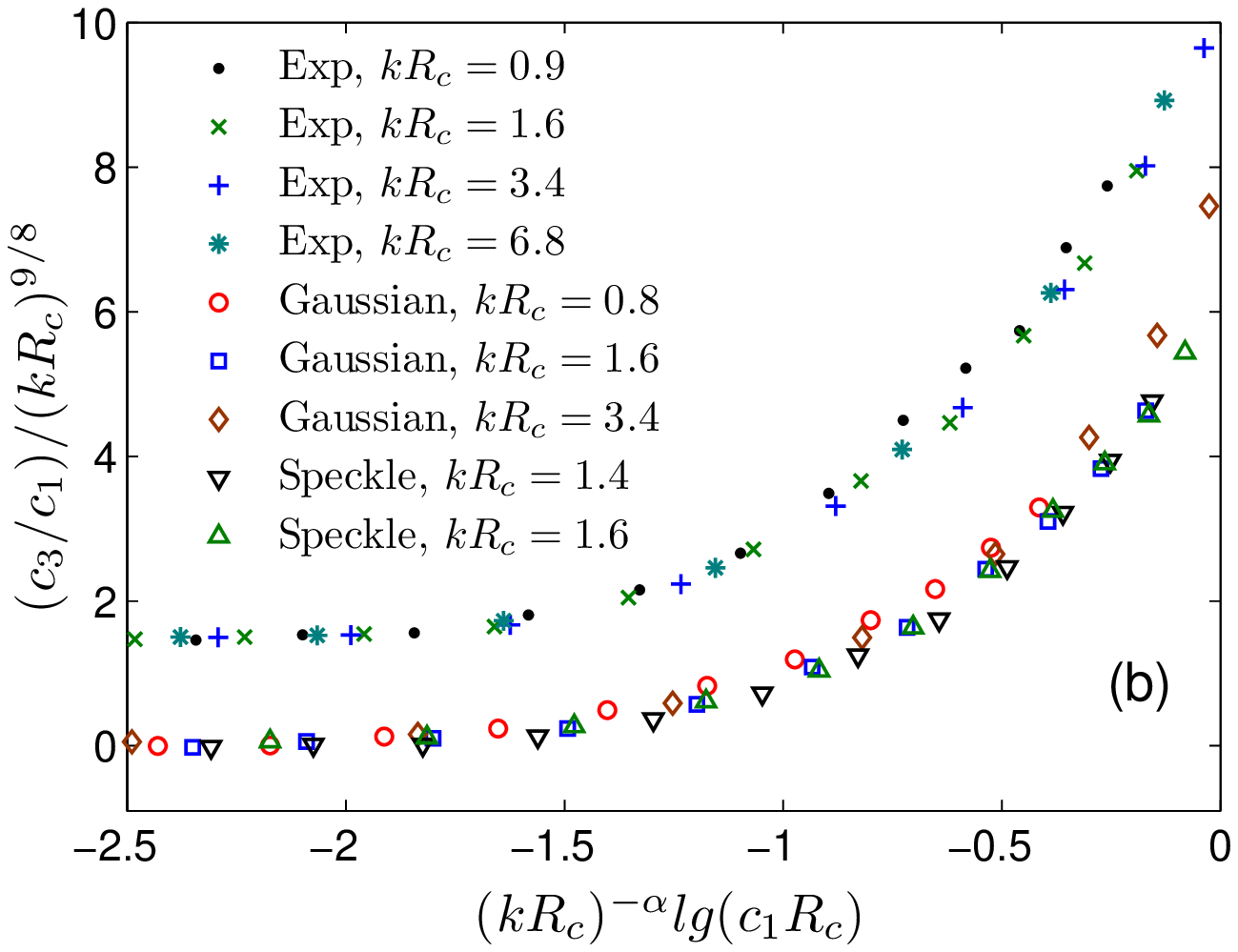}%
}%
\end{tabular}%
\caption{(color online) Demonstration of the empirical scaling (\ref
{Empir_Scaling}) for the cumulant coeffcients $c_2$ (upper panel)
and $c_3$ (lower panel). Legends in (b) apply to both panels and specify type of correlation and value of $k R_c$.
We use $\alpha=0$ for the exponential correlation, and $\alpha
=0.4$ - for the Gaussian and the speckle correlations.
For better visibility, data for the exponential correlation are offset from zero
by a constant (in both panels). Inset: same as in the main panel (a), but plotted in a logarithmic scale for
$y$-axis and without the offset.
}%
\label{Fig. Scaling_1}%
\end{figure}%

\subsection{Scaling of the cumulant ratios $c_{n}/c_{1}$}

In sufficiently strong disorder, such that $c_{1}R_{c}\gg1$, cumulant ratios
$c_{n}/c_{1}$ are described by the semiclassical relations
(\ref{Q-Cl_scaling_gen}). To verify these relations, in Fig.
\ref{Fig. Scaling_2} we plot $\left(  c_{n}/c_{1}\right)  /\left(  c_{1}%
R_{c}\right)  ^{n-1}$ for $n=2$,$3$ as a function of $\epsilon/V_{0}$ and
compare it to the semiclassical prediction $S_{n-1}B_{n}\left(  \frac
{\epsilon}{V_{0}}\right)  f_{n}\left(  \frac{\epsilon}{V_{0}}\right)  $,
according to Eq. (\ref{Q-Cl_scaling_gen}). For Gaussian disorder, functions
$f_{n}\left(  \frac{\epsilon}{V_{0}}\right)  $ are given by Eq.
(\ref{Q-Cl_f_n(App)}), and we use values $B_{2}=0.8$ and $B_{3}=1.2$, as
explained in text after Eq. (\ref{Q-Cl_c_2&3}). Fig. \ref{Fig. Scaling_2}
shows that the obtained numerical values converge to the theoretical curve for
$\epsilon/V_{0}<1$. As expected, the semiclassical approximation improves for
smaller $\epsilon/V_{0}$ and larger $R_{c}$. Let us note, for clarity, that in
our presentation we increase the disorder amplitude $V_{0}$ keeping fixed
$\epsilon>0$ and fixed $R_{c}$. In this parametrization, cumulant ratios
$c_{n}/c_{1}$ grow monotonically with $\beta$, roughly as $\left(  c_{1}%
R_{c}\right)  ^{n-1}$ (cf. Fig. \ref{Fig. Res_cums}). If, however, relative
disorder strength is increased by decreasing $\epsilon$ at fixed $V_{0}$ and
$R_{c}$, then ratios $c_{n}/c_{1}$ would eventually vanish for sufficiently
large and negative $\epsilon$, as can be seen from Eq. (\ref{Q-Cl_scaling_gen}%
) [because $V_{0}/\left\vert \epsilon\right\vert $ becomes small in this limit].

Semiclassical scaling (\ref{Q-Cl_scaling_gen}) is not applicable in weak and
moderate disorder. In this regime, and for $kR_{c}\gtrsim1$, we find
empirically that deviation from the weak disorder relations (\ref{SPS rel})
for $n=2,3$ is described by the following approximate scaling%
\begin{equation}
\frac{c_{n}}{c_{1}}-\delta_{2,n}=\left(  kR_{c}\right)  ^{\nu_{n}}\mu
_{n}\left(  \frac{\ln c_{1}R_{c}}{\left[  kR_{c}\right]  ^{\alpha}}\right)  ,
\label{Empir_Scaling}%
\end{equation}
where $\delta_{n,2}$ is the Kronecker delta. Here $\mu_{n}\left(  x\right)  $
are some dimensionless monotonically increasing functions, whose form depend
on the type of correlation. Exponent $\alpha$ is not universal as well, and we
find that $\alpha\approx0$, for the exponential, and $\alpha\approx0.4$ for
the Gaussian and the speckle correlation functions (Table \ref{Table}). On the
contrary, in the considered Gaussian models, we obtain $\nu_{2}\approx3/4$ and
$\nu_{3}\approx9/8$, irrespectively of the disorder correlation. In Fig.
\ref{Fig. Scaling_1} we plot $\left(  c_{2}/c_{1}-1\right)  /\left(
kR_{c}\right)  ^{3/4}$ and $\left(  c_{3}/c_{1}\right)  /\left(
kR_{c}\right)  ^{9/8}$ against $\left(  kR_{c}\right)  ^{-\alpha}\lg
c_{1}R_{c}$, which demonstrates that scaling (\ref{Empir_Scaling}) holds for a
rather broad range of the considered values of $kR_{c}$ (see legends) as long
as $c_{1}R_{c}<1$. A logarithmic scale plot of $\left(  c_{2}/c_{1}-1\right)
/\left(  kR_{c}\right)  ^{3/4}$, given in the inset of the upper panel, shows
that this scaling holds also at small values of $c_{1}R_{c}$ (the same is true
also for $c_{3}/c_{1}$).

Let us stress that relations (\ref{Empir_Scaling}) become meaningless in the
white noise limit $kR_{c}\rightarrow0$, and are applicable only for
$kR_{c}\gtrsim1$, when deviation from the weak disorder relations
(\ref{SPS rel}) is dominated by the effects of the correlations. According to
Eq. (\ref{Empir_Scaling}), the latter is controlled by the ratio of the
correlation to the localization lengths, $c_{1}R_{c}$.

As seen in Fig. (\ref{Fig. Scaling_1}), functions $\mu_{n}$ are rather
similar, though not identical, for different models of correlation. The major
distinction in a specific form of relation (\ref{Empir_Scaling}) for different
(Gaussian) models appears in the value of the exponent $\alpha$. It is
interesting to note that both $\mu_{n}$ and $\alpha$ practically coincide for
the models with Gaussian and speckle correlation functions. This can be
related to the qualitative similarity of their power spectra, given by the
Gaussian and the "tent" function respectively, which have either effective or
exact cutoff of the order of $R_{c}^{-1}$ (the "tent" function, $\left(
1-\left\vert x\right\vert \right)  \Theta\left(  1-\left\vert x\right\vert
\right)  $, was introduced in Sec. \ref{Sect: num tests}). On the contrary,
power spectrum of the exponentially correlated disorder is given by the slowly
decaying Lorentzian function. Note that disorder power spectrum appears, e.g.,
in the Born approximation (\ref{Born approx}) for $\Lambda$ (and LE $c_{1}$),
which shows close relation between the properties of localization and those of
the disorder power spectrum.
\begin{figure}
[ptb]
\begin{center}
\includegraphics[
height=2.2939in,
width=3.2146in
]%
{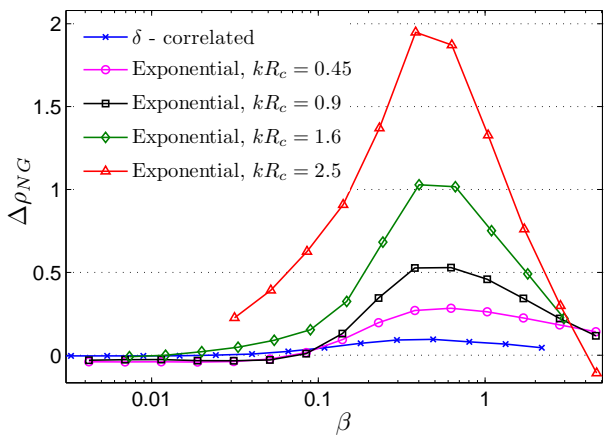}%
\caption{(color online) The non-Gaussian part of $\rho$, $\Delta\rho_{NG}%
=\rho-\rho_{G}$, where $\rho_{G}=\left(  1+c_{2}/c_{1}\right)  /2$, as a
function of $\beta$. The shown data correspond to the $\delta$-correlated and
exponentially correlated disorder with different values of $kR_{c}$, as
indicated in the legends. $\Lambda$ is calculated analytically using the RWNA
method. As noted in Sec. \ref{Sect: num tests}, approximation error of this
method becomes significant for $kR_{c}>1$. In the presented figure, this does
results in false values for $\beta\lesssim0.1$ (simulated values are much
closer to zero). However, the approxiamtion error is not critical in the
region of main interest, $0.1\lesssim\beta\lesssim1$, where it tends to
underestimate, rather than overestimate, the actual values of $\Delta\rho
_{NG}$. Thus, the presented effect is not an artifact of the approximation.}%
\label{Fig. del_rho}%
\end{center}
\end{figure}

\subsection{Extreme fluctuation of $T^{-1}$\label{Sect: Extreme_fluct}}

Beyond weak disorder, LE ratio $\rho$ increases with the correlation parameter
$kR_{c}$ and, as a function of the disorder intensity, exhibits a peak
\cite{Note-peak-of-rho} at moderate disorder (Fig. \ref{Fig. Res_gle}). Thus,
as expressed in terms of the ratio between the mean and the typical values of
the inverse transmission coefficient $T^{-1}$, Eq.
(\ref{Extreme_fluct_measure}), also the extreme relative fluctuation of
$T^{-1}$ is peaked at moderate disorder and is enhanced by the disorder
correlations. This conclusion is consistent with our observation that
statistical convergence of the Monte-Carlo simulation of the generalized LE
$\Lambda$ is most slow at moderate disorder and for larger values of $R_{c}$,
as indicated by the error bars in Figs. \ref{Fig. 1} and \ref{Fig. Speckle}.
Note that relative fluctuation (\ref{Extreme_fluct_measure}) depends
exponentially on $\rho l$, where $l\equiv c_{1}L$. Therefore, for
$kR_{c}\gtrsim1$ and $l>1$, correlations lead to the exponentially large
enhancement of the extreme fluctuation of $T^{-1}$ as compared to the case of
the $\delta$-correlated disorder.

It is instructive to compare $\rho$ to its Gaussian part $\rho_{G}=\frac{1}%
{2}\left(  1+\frac{c_{2}}{c_{1}}\right)  $. The latter is obtained by
discarding the non-Gaussian terms, $n\geq3$, in Eq. (\ref{Ro_def}), which is
equivalent to approximating the asymptotic distribution of $\ln T$ by the
Gaussian one with the same mean and variance, equal to $2c_{1}L$ and $4c_{2}L$
respectively. Correspondingly, contribution of the non-Gaussian corrections is
represented by the difference $\Delta\rho_{NG}\equiv\rho-\rho_{G}=\frac{1}%
{4}\sum_{n=3}^{\infty}\frac{2^{n}}{n!}\frac{c_{n}}{c_{1}}$. In Fig.
\ref{Fig. del_rho}\ we plot $\Delta\rho_{NG}$ for the $\delta$-correlated and
the exponentially correlated disorder (other models of correlation exhibit
similar behavior). In agreement with relations (\ref{SPS rel}), $\Delta
\rho_{NG}$ vanishes in the weak disorder limit. Beyond this regime,
$\Delta\rho_{NG}$ is positively peaked at moderate disorder, and the magnitude
of the peak increases with $kR_{c}$. Positive values of $\Delta\rho_{NG}$ mean
that the low-$T$ tail of the $\ln T$-distribution is heavier than in the
Gaussian approximation in a sense that $\left\langle T^{-1}\right\rangle $
and, thus, the relative fluctuation (\ref{Extreme_fluct_measure}), are larger
than in the corresponding Gaussian distribution. The result in Fig.
\ref{Fig. del_rho} shows that this "super-Gaussian" effect is enhanced by the
disorder correlations and is most prominent at moderate disorder. Let us
stress, however, that this statement applies only to the integral quantity
(\ref{Extreme_fluct_measure}) and does not imply any specific form (e.g.,
super- or sub-Gaussian) for the asymptotic decay law of the low-$T$ tail in
the $\ln T$-distribution. The latter was studied , e.g., in Refs.
\cite{Pikovsky-03}, \cite{Opt-Fluct}, \cite{Raikh} using different methods.

\section{Conclusions\label{Sect: Coclus}}

We have considered statistical properties of the transmission coefficient $T$
of a one-dimensional disordered system described by the Schr\"{o}dinger Eq.
(\ref{Schrodinger}) with a Gaussian correlated disorder. The main focus of our
study was an effect of the correlations on the transmission distribution,
which was characterized in terms of the dimensionless ratio between the
generalized and the usual Lyapunov exponents, $\rho=\Lambda/c_{1}$, as well as
ratios $c_{2}/c_{1}$ and $c_{3}/c_{1}$ of the asymptotic cumulants of $\ln
T^{-1}$. Both analytical and numerical methods were employed for calculation
of $\Lambda$ and $c_{n}$.

First, a small-$R_{c}$ approximation was developed for the generalized LE
$\Lambda$, which is not limited to weak disorder and also is able to account
for the non-trivial effects of the correlations. To this end, we obtained an
infinite hierarchy of integro-differential equations on the second moments of
the wave function and its functional derivatives with respect to the disorder.
We have shown that this hierarchical chain can be truncated starting from the
second level to obtain a non-trivial approximation, which accounts for the
correlations of the disorder. Note that termination of this hierarchy on the
first level accounts only for the intensity of the noise, while all
information on the correlation properties is lost. This non-trivial closure
allowed us to obtain an analytical solution for the generalized LE in an
implicit form as a largest real root of a non-linear algebraic equation. The
obtained approximation is valid formally for $kR_{c}\ll1$ and $\Lambda
R_{c}\ll1$, where $k=\sqrt{\left\vert \epsilon\right\vert }$ is the wave
number. It turned out that a standard white noise approximation
\cite{Klatzkin} does not allow to treat the generalized LE in the case of
sub-exponentially decaying correlations. To overcome this obstacle, we have
proposed a simple self-consistent regularization, which extends the
applicability of the approximation to any correlation function satisfying
$\left\vert \int_{0}^{\infty}C_{2}\left(  x\right)  dx\right\vert <\infty$.
The derived small-$R_{c}$ approximation was compared to numerical simulations
and a good agreement was found.

The asymptotic cumulants of $\ln T$ were calculated analytically within the
semiclassical approximation applicable in a special case of strong disorder,
$\epsilon\,<V_{0}$ and $R_{c}\sqrt{V_{0}-\epsilon}\gg1$. For an arbitrary
strength of disorder, cumulant coefficients $c_{1,2,3}$ were obtained by
numerical simulations.

Using these analytical and numerical methods, the generalized LE and the first
three cumulants of $\ln T$ were calculated for several models of correlations
and for different values of $R_{c}$, which enabled us to investigate effects
of correlations on the form of the transmission distribution. In order to
study transition between the regimes of weak and strong disorder, we have
considered only positive energies $\epsilon$, where it was convenient to
introduce the dimensionless disorder intensity $\beta$, defined in Eq.
(\ref{beta-def}).

In sufficiently weak disorder, we obtained $c_{2}/c_{1}=1$, $c_{3}/c_{1}=0$
and $\rho=1$ in all the considered cases. Thus, as expected from previous work
(e.g., Refs. \cite{Tamura-93},\cite{Petri-96-97},\cite{Zaslavsky-97}%
,\cite{Deych-03}), in weak disorder, correlations do not destroy SPS and do
not modify the universal relations (\ref{SPS rel}). In the white noise limit,
$kR_{c}\ll1$, this regime is realized for $c_{1}/k\sim\beta\ll1$. For
$kR_{c}\gtrsim1$, the relevant control parameter is the ratio of the
correlation to the localization lengths, $c_{1}R_{c}$. Quantity $\left(
c_{1}R_{c}\right)  ^{-1}$ is interpreted naturally as a measure of
randomization of the disorder potential on the scale of the localization
length. Correspondingly, the weak disorder universality, expressed by
relations (\ref{SPS rel}), takes place when $c_{1}R_{c}\ll1$ (for
$kR_{c}\gtrsim1$, this means that disorder is weak also in the conventional
sense $c_{1}/k\ll1$).

Relations (\ref{SPS rel}) are not valid beyond weak disorder. In the white
noise model, the corresponding deviation from these relations is weak for any
strength of the disorder (for $\epsilon\gtrsim0$), in a sense that values of
$\rho$ and $c_{2}/c_{1}$ stay very close to unity, while $c_{3}/c_{1}$ remains
small compared to unity. On the contrary, in correlated disorder, these
quantities depend strongly on the strength of the disorder through the control
parameter $c_{1}R_{c}$. Regarding this parametric dependence, we discuss two
regimes: $c_{1}R_{c}\lesssim1$ and $c_{1}R_{c}\gg1$.

The first regime, $c_{1}R_{c}\lesssim1$, corresponds to transition from weak
to moderate disorder. In this regime and for $kR_{c}\gtrsim1$, cumulant ratios
$c_{2}/c_{1}$ and $c_{3}/c_{1}$\ are described by the approximate scaling
relations (\ref{Empir_Scaling}) with parameters $kR_{c}$ and $c_{1}R_{c}$.
This relations demonstrate that, starting from weak disorder ($c_{1}R_{c}\ll
1$), ratios $c_{n}/c_{1}$ increase gradually with parameter $c_{1}R_{c}$ and,
for $c_{1}R_{c}\sim1$, arrive at values which can be much larger than unity
(depending on $kR_{c}$). In such a case, the bulk of the $\ln T$-distribution
becomes much broader than in weak or in the white noise disorder.

While the small-$n$ cumulant coefficients $c_{n}$ describe bulk of the $\ln
T$-distribution, LE ratio $\rho$ is a measure of the extreme relative
fluctuation of $T^{-1}$, expressed by the ratio between the mean and the
typical values of $T^{-1}$, Eq. (\ref{Extreme_fluct_measure}). Like the
cumulant ratios, $\rho$ increases from weak to moderate disorder and its
deviation from the weak disorder value $\rho=1$ is strongly enhanced by the
disorder correlations (namely, exceeds unity for $kR_{c}\gtrsim1$). As a
function of the disorder strength, $\rho$ and, thus, the extreme relative
fluctuation (\ref{Extreme_fluct_measure}), are peaked at moderate disorder
(near $c_{1}R_{c}\sim0.1$). This peak of $\rho$ is associated with the
non-Gaussian corrections to the low-$T$ tail of the $\ln T$-distribution,
whose contribution to $\rho$ is positively peaked, i.e. "super-Gaussian", at
moderate disorder. The latter has the following simple interpretation in terms
of the disorder statistics. In moderate disorder, when energy $\epsilon$ is of
the order of $V_{0}$, wave propagation becomes affected by the under-barrier
tunneling through the rare but large peaks of the random potential.
For$\sqrt{V_{0}}R_{c}\sim kR_{c}\gtrsim1$, already a single large barrier
becomes a strong scatterer. Therefore, fluctuations in height and in
occurrence of these rare peaks lead to extreme deviations of $T^{-1}$ from its
typical value. In stronger disorder, the typical value grows significantly,
and relative contribution of the large rare barriers becomes less pronounced.
This simple explanation can be confirmed by calculating $\rho$ in the
framework of the semiclassical approximation (\ref{QuasiCl approx}) [not
presented here].

The second regime, $c_{1}R_{c}\gg1$, is realized when $\epsilon<V_{0}$ and
$R_{c}\sqrt{V_{0}-\epsilon}\gg1$. In these conditions, interference effects
are suppressed and localization is dominated by under-barrier tunneling. Then,
according to Eq. (\ref{QuasiCl approx}), the transmission distribution is
directly related to the statistics of the "excursions" of the random potential
above the level $V\left(  x\right)  =\epsilon$, and the cumulants of $\ln
T^{-1}$ satisfy the "semiclassical" scaling relations (\ref{Q-Cl_scaling_gen}%
), which were verified in numerical simulations. According to Eq.
(\ref{Q-Cl_scaling_gen}), the effect of correlations is expressed by the
simple relation $c_{n}/c_{1}\sim f_{n}\left(  \epsilon/V_{0}\right)  \left(
c_{1}R_{c}\right)  ^{n-1}$, where the coefficient $f_{n}\left(  \epsilon
/V_{0}\right)  \equiv\left\langle \left\langle \kappa^{n}\right\rangle
\right\rangle /\left\langle \kappa\right\rangle ^{n}$ depends only on the
one-point distribution of disorder. Semiclassical approximation can also be
used to calculate the generalized LE $\Lambda$. In particular, for $\left\vert
\epsilon\right\vert \ll V_{0}$ and $\sqrt{V_{0}}R_{c}\gg1$, we obtain scaling
$\rho\sim\left(  c_{1}R_{c}\right)  ^{1/3}$, which is specific to Gaussian
statistcs of disorder.

\begin{acknowledgments}
We are grateful to B. Shapiro for useful discussions. This work was supported
by the Israel Science Foundation under the grants No. 1067/06 and 1299/07.
\end{acknowledgments}

\appendix

\section{\label{Sect: GLE for WN}Exact solution for $\Lambda$ in $\delta
$-correlated disorder}

Equation (\ref{Evolut for <Y> with N-F}) decouples and can be solved exactly
for the $\delta$-correlated disorder (see e.g. Ref. \cite{Pikovsky-03}).
Namely, substituting $C_{2}\left(  t-\tau\right)  =g\delta\left(
t-\tau\right)  $ into (\ref{Evolut for <Y> with N-F}) and using the initial
condition $\left\langle \frac{\delta\mathbf{Y}\left(  t^{+}\right)  }{\delta
V\left(  t^{-}\right)  }\right\rangle =\mathcal{D}\left\langle \mathbf{Y}%
\left(  t\right)  \right\rangle $ (\ref{I.C. for Evolut for <dY/dh> with N-F}%
), one obtains
\begin{equation}
\partial_{t}\left\langle \mathbf{Y}\left(  t\right)  \right\rangle
=\mathcal{M}\left(  \epsilon,g\right)  \left\langle \mathbf{Y}\left(
t\right)  \right\rangle , \label{Evolut for <Y> in WN}%
\end{equation}
where%
\begin{equation}
\mathcal{M}\left(  \epsilon,g\right)  =\mathcal{C}+\frac{g}{2}\mathcal{D}%
^{2}=\sqrt{2}%
\begin{bmatrix}
0 & 1 & 0\\
-\epsilon & 0 & 1\\
g/\sqrt{2} & -\epsilon & 0
\end{bmatrix}
, \label{M-matrix}%
\end{equation}
and matrices $\mathcal{C}$ and $\mathcal{D}$ are defined in (\ref{C and D}).
Solution of Eq. (\ref{Evolut for <Y> in WN}) is%
\begin{equation}
\left\langle \mathbf{Y}\left(  t\right)  \right\rangle =e^{\mathcal{M}\left(
\epsilon,g\right)  t}\mathbf{Y}_{0}, \label{Solut for <Y> in WN}%
\end{equation}
where $\mathbf{Y}_{0}=\mathbf{Y}\left(  0\right)  $ is an initial condition.
The generalized LE $\Lambda$, Eq. (\ref{eq. generalized LE}), is given by the
largest real eigenvalue of the matrix $\frac{1}{4}\mathcal{M}\left(
\epsilon,g\right)  $. It is found from the cubic equation%
\begin{equation}
\Lambda\left(  4\Lambda^{2}+\epsilon\right)  =\frac{g}{8},
\label{GLE in WN implicite}%
\end{equation}
which has the following roots%
\begin{equation}
\Lambda_{1}=\frac{\rho\left(  g,\epsilon\right)  }{4}-\frac{\epsilon}%
{3\rho\left(  g,\epsilon\right)  },\ \Lambda_{2,3}=\frac{-\Lambda_{1}\pm
i\sqrt{3\Lambda_{1}^{2}+\epsilon}}{2}, \label{Lambda_i}%
\end{equation}
where $\rho\left(  g,\epsilon\right)  =\left(  g+\sqrt{g^{2}+\left(  \frac
{4}{3}\epsilon\right)  ^{3}}\right)  ^{1/3}$. The generalized LE $\Lambda$ is
given by $\Lambda_{1}$, which is real and positive. Other two eigenvalues are
complex for $\epsilon>-3\Lambda_{1}^{2}$ (i.e. $\epsilon>-\frac{3}{4}g^{2/3}%
$), and real otherwise. Matrix $\mathcal{M}\left(  \epsilon,\tilde{g}\right)
$ is not normal ($\mathcal{MM}^{\dagger}\neq\mathcal{M}^{\dagger}\mathcal{M}%
$), and its left and right eigenvectors, corresponding to the eigenvalues
$\Lambda_{i}$ in Eq. (\ref{Lambda_i}), are
\begin{align}
u_{i}^{R}  &  =\frac{1}{24\Lambda_{i}^{2}+2\epsilon}%
\begin{bmatrix}
1, & 2\sqrt{2}\Lambda_{i}, & \epsilon+8\Lambda_{i}^{2}%
\end{bmatrix}
^{T},\nonumber\\
u_{i}^{L}  &  =%
\begin{bmatrix}
\epsilon+8\Lambda_{i}^{2}{}, & 2\sqrt{2}\Lambda_{i}, & 1
\end{bmatrix}
^{\dagger}. \label{Left+Right EigVecs WN}%
\end{align}
These eigenvectors satisfy the normalization $\left(  u_{i}^{L},u_{j}%
^{R}\right)  =\delta_{ij}$.

\section{Born approximation for $\Lambda$\label{Sect: Born Approx}}

The systematic weak disorder expansion (i.e. in powers of the disorder
amplitude $V_{0}$) for the generalized LE $\Lambda$ was considered in Ref.
\cite{Tessieri-02}. Here we only note that the Born approximation\ for
$\Lambda$ can be obtained by substitution of the zero order solution for
$\left\langle \frac{\delta\mathbf{Y}\left(  t\right)  }{\delta V\left(
\tau\right)  }\right\rangle $ into (\ref{Evolut for <Y> with N-F}). The zero
order solution for $\left\langle \frac{\delta\mathbf{Y}\left(  t\right)
}{\delta V\left(  \tau\right)  }\right\rangle $ is obtained by neglecting the
noise term in equation (\ref{Evolut for <dY/dh> with N-F}):
\[
\frac{\partial}{\partial t}\left\langle \frac{\delta\mathbf{Y}\left(
t\right)  }{\delta V\left(  \tau\right)  }\right\rangle =\mathcal{C}%
\left\langle \frac{\delta\mathbf{Y}\left(  t\right)  }{\delta V\left(
\tau\right)  }\right\rangle ,\ \left\langle \frac{\delta\mathbf{Y}\left(
\tau^{+}\right)  }{\delta V\left(  \tau\right)  }\right\rangle =\mathcal{D}%
\left\langle \mathbf{Y}\left(  \tau\right)  \right\rangle ,
\]
which has the solution%
\begin{equation}
\left\langle \frac{\delta\mathbf{Y}\left(  t\right)  }{\delta V\left(
\tau\right)  }\right\rangle =e^{\mathcal{C}\left(  t-\tau\right)  }%
\mathcal{D}\left\langle \mathbf{Y}\left(  \tau\right)  \right\rangle .
\label{Sol dY/dh free}%
\end{equation}
Inserting this solution into Eq. (\ref{Evolut for <Y> with N-F}) yields an
equation closed with respect to $\left\langle \mathbf{Y}\left(  t\right)
\right\rangle $:%
\begin{equation}
\partial_{t}\left\langle \mathbf{Y}\right\rangle =\mathcal{C}\left\langle
\mathbf{Y}\right\rangle +\int_{0}^{t}C_{2}\left(  t-\tau\right)
\mathcal{D}e^{\mathcal{C}\left(  t-\tau\right)  }\mathcal{D}\left\langle
\mathbf{Y}\left(  \tau\right)  \right\rangle d\tau,
\label{Evolut for <Y> in Born}%
\end{equation}
where pair correlation function $C_{2}\left(  t-\tau\right)  $ is defined in
Eq. (\ref{eq. DO properties}). Then, lowest order calculation of the
asymptotic growth rate of the solution $\left\langle \mathbf{Y}\left(
t\right)  \right\rangle $ yields the Born approximation%
\begin{equation}
\frac{\Lambda}{k}=\frac{g}{8\epsilon^{3/2}}\int_{0}^{\infty}ds\Gamma\left(
s\right)  \cos\left(  2kR_{c}s\right)  +o\left(  \frac{g}{8\epsilon^{3/2}%
}\right)  , \label{Born approx}%
\end{equation}
where $k=\sqrt{\epsilon}$ and $g=2V_{0}^{2}R_{c}$. This expression coincides
with the Born approximation for LE $c_{1}$ \cite{LGP-Introduction}.

\section{Method of numerical simulations\label{Sect: Numrics}}

The numerical simulations were performed using the tight-binding (TB) model%
\begin{equation}
E\psi_{n}=\varepsilon_{n}\psi_{n}-\psi_{n+1}-\psi_{n-1},
\label{Tight-binding Shrod. Eq.}%
\end{equation}
with energy near the band edge, $0<E+2\ll1$, and the diagonal disorder
$\left\vert \varepsilon_{n}\right\vert \lesssim E+2$. In this regime, Eq.
(\ref{Tight-binding Shrod. Eq.}) is a good approximation to the continuous
model (\ref{eqn1}). We relate the continuous model with energy $\epsilon
=k^{2}$ and potential $V\left(  x\right)  $ to the TB counterpart by setting
\begin{equation}
E=-2\cos k,\quad\varepsilon_{n}=\frac{\sin k}{k}\left.  V\left(  x\right)
\right\vert _{x=n}. \label{Relation btw DO in TB and Cont. Model}%
\end{equation}
The correlated Gaussian disorder $\varepsilon_{n}$ was generated by filtering
sequences of independent Gaussian random variables. The real space convolution
with a proper kernel was used to obtain the "short-range" exponential
($\Gamma\left(  x\right)  =e^{-x}$) and Gaussian ($\Gamma\left(  x\right)
=e^{-\pi x^{2}/4}$) correlations, while the Fourier space filter was applied
to obtain the power law correlation $\Gamma\left(  x\right)
=\operatorname{sinc}^{2}\frac{x}{2}$.

In all simulations presented in this paper, we have fixed the energy at
$E=-1.95$ and varied values of the disorder amplitude $V_{0}$ and the
correlation radius $R_{c}$. Note that for $E=-1.95$, the corresponding
wavelength of the solution for a pure system ($\varepsilon_{n}=0$) is equal to
about $28$ sites. Therefore, approximation to a continuum is good as long as
$V_{0}\lesssim E$.

A standard transfer matrix formalism (see e.g. Ref. \cite{Liu-86}) was used,
in which the TB equation (\ref{Tight-binding Shrod. Eq.}) is rewritten in the
matrix form%
\begin{equation}
\left(
\begin{array}
[c]{c}%
\psi_{n+1}\\
\psi_{n}%
\end{array}
\right)  =\mathcal{T}_{n}\left(
\begin{array}
[c]{c}%
\psi_{n}\\
\psi_{n-1}%
\end{array}
\right)  ,\ \mathcal{T}_{n}=\left[
\begin{array}
[c]{cc}%
\left(  \varepsilon_{n}-E\right)  & -1\\
1 & 0
\end{array}
\right]  , \label{Single-site transf. Matr}%
\end{equation}
where $\mathcal{T}_{n}$ is a single-site transfer matrix. Then, solution of
the initial value problem is given by%
\begin{equation}
\left(
\begin{array}
[c]{c}%
\psi_{N+1}\\
\psi_{N}%
\end{array}
\right)  =\mathcal{T}_{N,1}\left(
\begin{array}
[c]{c}%
\psi_{1}\\
\psi_{0}%
\end{array}
\right)  ,\quad\mathcal{T}_{N,1}\mathbf{=}\mathcal{T}_{N}...\mathcal{T}%
_{2}\mathcal{T}_{1}, \label{Transfer Matr Eq.}%
\end{equation}
where $\mathcal{T}_{1,N}$ is the total transfer matrix for the system of
length $N$.

In analogy with Eq. (\ref{Y_def}), one can introduce vector $\mathbf{Y}%
_{n}=(u_{n}^{2},\sqrt{2}u_{n}v_{n},v_{n}^{2})^{T}$, where $u_{n}\equiv\psi
_{n}$ and $v_{n}=\left(  \psi_{n}-\psi_{n-1}\right)  /k\approx k^{-1}%
\partial_{x}\psi$. Similarly to (\ref{Transfer Matr Eq.}), the solution for
$\mathbf{Y}_{N}$ can be write as%
\begin{equation}
\mathbf{Y}_{N+1}=\mathcal{K}_{N,1}\mathbf{Y}_{1},
\end{equation}
where $\mathbf{Y}_{1}$ is an initial condition, and $\mathcal{K}_{N,1}$ is the
transfer matrix for $\mathbf{Y}_{N}$, which is readily expressed in terms of
the elements of $\mathcal{T}_{N,1}$. The largest eigenvalue of $\mathcal{K}%
_{N,1}$ is equal to the square modulus of the largest eigenvalue of
$\mathcal{T}_{N,1}$. Finally, ensemble average over the disorder realizations
yields%
\[
\left\langle \mathbf{Y}_{N+1}\right\rangle =\left\langle \mathcal{K}%
_{N,1}\right\rangle \mathbf{Y}_{1}.
\]

Cumulant coefficients $c_{n}$, defined in Eq. (\ref{Cn coefs-def}), are given
by the asymptotic linear growth rate of the cumulants $2^{-n}\left\langle
\left\langle \ln^{n}\kappa_{N}\right\rangle \right\rangle $ with the system
length $N$, where $\kappa_{N}$ denotes the largest eigenvalue of the matrix
$\mathcal{K}_{1,N}$. The asymptotic slope was calculated by the linear fit,
which have to exclude the region of the initial transient of the order of a
few localization lengths. The ensemble average was performed over $\sim
10^{6}\div10^{7}$ realizations of disorder.

The generalized Lyapunov exponent $\Lambda$, Eq. (\ref{eq. generalized LE}),
was calculated as a linear slope of $\frac{1}{4}\ln\left\langle \kappa
_{N}\right\rangle $. Alternatively, $\Lambda$ could be found as a slope of the
logarithm of the largest eigenvalue of $\left\langle \mathcal{K}%
_{N,1}\right\rangle $, which gives practically the same result. About $10^{8}$
realization were generated to calculate each value of $\left\langle \kappa
_{N}\right\rangle $.

Monte-Carlo simulation of the generalized LE can be a quite challenging task,
as is briefly explained in the following. In numerical simulation of the
generalized LE one have to deal with two restrictions on the system size, both
from below and from above. The lower bound is determined by the width of the
transient to the asymptotic behavior%
\begin{equation}
\ln\left\langle \kappa_{N}\right\rangle =N\Lambda+const. \label{kap_N-asympt}%
\end{equation}
Width of this transient is at least of the order of $R_{c}$. This follows from
the form of the differential equation (\ref{Evolut for <Y> with N-F}) for
$\left\langle Y\left(  t\right)  \right\rangle $, which suggests that growth
rate of $\left\langle Y\left(  t\right)  \right\rangle $ can not stabilize
unless $t\gg R_{c}$. This is because the correlation function in the integral
on the right hand side of Eq. (\ref{Evolut for <Y> with N-F}) decays on the
scale of $R_{c}$. In the white noise limit, $R_{c}\rightarrow0$, the transient
region is absent, as follows from the exact analytical result (Appendix
\ref{Sect: GLE for WN}) and was observed numerically. Therefore, we assume
that the width of the transient is of the order of $R_{c}$, and other scales,
such as the localization length, are less important (unlike the case of
$c_{n}$, $n>2$).

The upper bound on the system size is determined empirically from the
numerical data as a value of $N$, beyond which the linear dependence in Eq.
(\ref{kap_N-asympt}) is violated. This computational artifact originates from
the insufficient statistics in averaging of the broadly distributed quantity
$\kappa_{N}$, which is equivalent to $T^{-1}$. According to Eqs.
(\ref{Cn coefs-def_0}) and (\ref{SPS rel}), distribution of $\kappa_{N}$ is
log-normal in weak disorder, while some corrections to the limiting log-normal
form appear in stronger disorder. As follows from Eqs. (\ref{T_cums_dimless})
and (\ref{Extreme_fluct_measure}), this distribution becomes increasingly
broad and heavy-tailed with the increase of the dimensionless system length
$l=c_{1}N$. For large $l$, long tails of the distribution, which dominate the
theoretical mean of $\kappa_{N}$, are typically under-sampled in simulations
with a finite number of realization. As a result, the obtained values of
$\ln\left\langle \kappa_{N}\right\rangle $ become typically underestimated
(formally, expectation value of $\ln\left\langle \kappa_{N}\right\rangle $
becomes smaller that $\ln\left\langle \kappa_{N}\right\rangle _{\infty}$,
where $\left\langle \kappa_{N}\right\rangle _{\infty}$ is the theoretical
mean). This effect increases with the system size, the relevant length scale
being the localization length $c_{1}^{-1}$. Therefore, the upper bound on the
system length is of the order of a few $c_{1}^{-1}$. Thus, the upper and the
lower bounds eventually coincide in sufficiently strong disorder, since
$c_{1}^{-1}$ becomes small. In such a case, numerical calculation of $\Lambda$
becomes impossible, unless the number of the realizations is increased
dramatically (for the exactly log-normal $\kappa_{N}$, it can be shown that
the improvement is logarithmically slow). In moderate disorder, the small
range between the lower and the upper bounds results in large uncertainty in
the calculated $\Lambda$, as indicated by the error bars in Figs. \ref{Fig. 1}
and \ref{Fig. Speckle}.

\section{Auxiliary formulas for semiclassical
approximation\label{Sect: Auxiliary}}

Cumulants $\left\langle \left\langle \kappa^{n}\right\rangle \right\rangle $
are easily written in terms of $\left\langle \left\langle \kappa
^{n}\right\rangle \right\rangle _{\epsilon}$ and $\left\langle \left\langle
\Theta_{\epsilon}^{n}\left(  V\right)  \right\rangle \right\rangle $ using
that $\Theta_{\epsilon}^{n}\left(  V\right)  =\Theta_{\epsilon}\left(
V\right)  $ and $\left\langle \kappa^{n}\right\rangle =\left\langle \kappa
^{n}\right\rangle _{\epsilon}\left\langle \Theta_{\epsilon}\left(  V\right)
\right\rangle $, where $\left\langle \cdots\right\rangle _{\epsilon}$ stands
for the average with the distribution $P_{\epsilon}\left(  V\right)
\equiv\Theta_{\epsilon}\left(  V\right)  P\left(  V\right)  /\int_{\epsilon
}^{\infty}dVP\left(  V\right)  $, introduced in Sec. \ref{Sect: Semiclass}.
For example, the second cumulant, $\left\langle \left\langle \kappa
^{2}\right\rangle \right\rangle \equiv\left\langle \kappa^{2}\right\rangle
-\left\langle \kappa\right\rangle ^{2}$, becomes%

\begin{equation}
\left\langle \left\langle \kappa^{2}\right\rangle \right\rangle =\left\langle
\kappa^{2}\right\rangle _{\epsilon}\left\langle \Theta_{\epsilon}\left(
V\right)  \right\rangle -\left\langle \kappa\right\rangle _{\epsilon}%
^{2}\left\langle \Theta_{\epsilon}\left(  V\right)  \right\rangle ^{2},
\end{equation}
which,\ using $\left\langle \left\langle \Theta_{\epsilon}^{2}\left(
V\right)  \right\rangle \right\rangle =\left\langle \Theta_{\epsilon}\left(
V\right)  \right\rangle -\left\langle \Theta_{\epsilon}\left(  V\right)
\right\rangle ^{2}$ and $\left\langle \left\langle \kappa^{2}\right\rangle
\right\rangle _{\epsilon}=\left\langle \kappa^{2}\right\rangle _{\epsilon
}-\left\langle \kappa\right\rangle _{\epsilon}^{2}$, yields after some
rearrangement%
\begin{equation}
\left\langle \left\langle \kappa^{2}\right\rangle \right\rangle =\left\langle
\left\langle \kappa^{2}\right\rangle \right\rangle _{\epsilon}\left\langle
\Theta_{\epsilon}\left(  V\right)  \right\rangle +\left\langle \kappa
\right\rangle _{\epsilon}^{2}\left\langle \left\langle \Theta_{\epsilon}%
^{2}\left(  V\right)  \right\rangle \right\rangle .
\end{equation}
Similar expansion for the third cumulant, $\left\langle \left\langle
\kappa^{3}\right\rangle \right\rangle =\left\langle \kappa^{3}\right\rangle
-3\left\langle \kappa^{2}\right\rangle \left\langle \kappa\right\rangle
+2\left\langle \kappa\right\rangle ^{3}$, gives expression (\ref{Q-Cl_c_2&3})
for $f_{3}=\left\langle \left\langle \kappa^{3}\right\rangle \right\rangle
/\left\langle \kappa\right\rangle ^{3}$.

For Gaussian distribution $P\left(  V\right)  =\left(  2\pi V_{0}^{2}\right)
^{-1/2}e^{-V^{2}/2V_{0}^{2}}$, definition (\ref{Q-Cl_c1}) yields the following
expression for LE:%
\begin{equation}
c_{1}=V_{0}^{1/2}2^{-9/4}e^{-r^{2}/2}U\left(  \frac{3}{4},\frac{1}{2}%
,\frac{r^{2}}{2}\right)  ,\quad r\equiv\frac{\epsilon}{V_{0}},
\label{Q-Cl_c1(App)}%
\end{equation}
while functions $f_{n}\left(  \frac{\epsilon}{V_{0}}\right)  \equiv
\left\langle \left\langle \kappa^{n}\right\rangle \right\rangle /\left\langle
\kappa\right\rangle ^{n}$ for $n=2,3$ are given by ($r\geq0$)%
\begin{align}
f_{2}\left(  r\right)   &  =-1+8\frac{2e^{r^{2}/2}-\sqrt{2\pi}re^{r^{2}%
}\operatorname{erfc}\left(  r/\sqrt{2}\right)  }{\sqrt{\pi}U^{2}\left(
\frac{3}{4},\frac{1}{2},\frac{r^{2}}{2}\right)  },\nonumber\\
f_{3}\left(  r\right)   &  =2-24\frac{2e^{\frac{r^{2}}{2}}-\sqrt{2\pi
}re^{r^{2}}\operatorname{erfc}\frac{r}{\sqrt{2}}}{\sqrt{\pi}U^{2}\left(
\frac{3}{4},\frac{1}{2},\frac{r^{2}}{2}\right)  }+\nonumber\\
&  +\frac{32r^{\frac{1}{2}}e^{\frac{5r^{2}}{4}}\left[  \left(  1+r^{2}\right)
K_{\frac{1}{4}}\left(  \frac{r^{2}}{4}\right)  -r^{2}K_{\frac{3}{4}}\left(
\frac{r^{2}}{4}\right)  \right]  }{2^{1/4}\sqrt{\pi}U^{3}\left(  \frac{3}%
{4},\frac{1}{2},\frac{r^{2}}{2}\right)  } \label{Q-Cl_f_n(App)}%
\end{align}
where $U\left(  a,b,z\right)  $ is a confluent hypergeometric function of the
second kind (or Tricomi function) and $K_{\nu}\left(  z\right)  $ is a
modified Bessel function of the second kind \cite{Abramovich}.


\begin{thebibliography}{99}
\bibitem {Note-T-phase-Coorel}Correlation between the transmission coefficient
and the reflection and transmission phases is not exactly zero, and it can not
be ignored in certain circumstances. One such example is calculation of the
transmission (conductance) distribution based on the phase formalism
\cite{Schomerus-03}. For our purposes, however, this correlation becomes
negligible in the limit of long system.
\bibitem {Note-peak-of-rho}According to our semiclassical treatment of strong
disorder (Sec. \ref{Sect: Semiclass}, Eq. (\ref{QC_GLE_res})), $\rho$ scales
as $\left(  \sqrt{V_{0}}R_{c}\right)  ^{1/3}$ for $\left\vert \epsilon
\right\vert \ll V_{0}$ and $\sqrt{V_{0}}R_{c}\gg1$. Thus, if $V_{0}$ is
increased with $\epsilon$ and $R_{c}$ fixed, then the local peak of $\rho$ at
moderate disorder is followed by subsequent increase at much stronger
disorder. The latter is not present in Fig. \ref{Fig. Res_gle}, which
corresponds to less strong disorder, $\sqrt{V_{0}}R_{c}\lesssim1$. In fact, we
checked that the non-monotonic dependence of $\rho$ on $V_{0}$ in moderate and
strong disorder can be obtained using the semiclassical approximation
(\ref{QuasiCl approx}).
\end{thebibliography}
\end{document}